\documentclass[12pt]{article}
\usepackage{epsfig}
\usepackage[T1]{fontenc}
\usepackage[latin2]{inputenc}
\usepackage{colordvi}
\usepackage{cite}
\def\bea{\begin{eqnarray}}
\def\eea{\end{eqnarray}}
\def\bq{\begin{quote}}
\def\eq{\end{quote}}

\def\gappeq{\mathrel{\rlap
{\raise.5ex\hbox{$>$}}
{\lower.5ex\hbox{$\sim$}}}}
\def\lappeq{\mathrel{\rlap{\raise.5ex\hbox{$<$}}
{\lower.5ex\hbox{$\sim$}}}}
\def\simlt{\stackrel{<}{{}_\sim}}

\newcommand{\beq}{\begin{equation}}
\newcommand{\eeq}{\end{equation}}

\def\varkappa{\omega}

\def\babar{\mbox{\sl B\hspace{-0.4em} {\scriptsize\sl A}\hspace{-0.37em} \sl
    B\hspace{-0.4em} {\scriptsize\sl A\hspace{-0.02em}R}}}

\newcounter{mnotecount}[section]

\begin{document}
\pagestyle{empty}
\begin{flushright}
IFT-2004/30\\
IPPP/04/85\\
DCPT/04/170\\
{\tt hep-ph/0412253}\\
{\bf \today}
\end{flushright}
\vspace*{5mm}
\begin{center}

{\large {\bf CP violation in $B^0_d\rightarrow\tau^+\tau^-$ decays}}\\
\vspace*{1cm}
{\bf Piotr~H.~Chankowski}$^1$, {\bf Jan~Kalinowski}$^{1,2}$, 
{\bf Zbigniew~W\c{a}s}$^3$\\ and {\bf Ma\l gorzata~Worek}$^{3,4}$\\
\vspace{0.3cm}

$^1$  Institute of Theoretical Physics, Warsaw University, Ho\.za 69, 00-681, 
Warsaw, Poland\\
$^2$ IPPP, University of Durham, South Road, Durham DH1 3LE, UK\\
$^3$ Institute of Nuclear Physics PAS, Radzikowskiego 152, 
31-3420, Cracow, Poland
\\
$^4$ Institute of Nuclear Physics, NCSR "Demokritos'', 15310 Athens, Greece
\\
\vspace*{1.7cm} 
{\bf Abstract} 
\end{center}
\vspace*{5mm}
\noindent
{
Establishing CP violation in $B^0(\bar{B}^0)\rightarrow l^+l^-$ decays 
requires a measurement of polarization of the final lepton pair, or a 
precise determination of the $B^0\to l^+l^-$ and $\bar B^0\to l^+l^-$ rates. 
We first  argue that if the amplitudes of these decays are dominated by 
the scalar and pseudoscalar Higgs penguin diagrams, as happens {\it e.g.} in 
supersymmetry with large $\tan\beta$, the CP asymmetries depend practically 
on only one CP violating phase. This phase can be large, of the order
of the CKM phase, leading to large CP asymmetries in the $\tau^+\tau^-$ decay 
channel of $B^0_d(\bar{B}^0_d)$ mesons, potentially measurable in BELLE or 
\babar\ experiments. Secondly, we show that the existing {\tt TAUOLA}
$\tau$-lepton decay library supplemented by its universal interface can 
efficiently be used to search for $B^0(\bar{B}^0)\rightarrow\tau^+\tau^-$ 
decays, and  to investigate how the CP asymmetry is reflected in realistic 
experimental observables.
}
\vspace*{1.0cm}
\date{\today} 


\rule[.1in]{14.5cm}{.002in}
\vspace*{0.2cm}

\newpage

\setcounter{page}{1}
\pagestyle{plain}

\section{Introduction}

40 years after its discovery in the Kaon system, CP violation has also been 
firmly established in $B$-physics in a series of high statistics experiments 
which enabled sufficiently precise measurements of the relevant observables. 
While the reported small difference between the time dependent CP asymmetries 
measured in $B\rightarrow J/\psi K_S$ and in $B\rightarrow\phi K_S$ decays 
\cite{differ}, if confirmed, would conflict with the Standard Model (SM) 
prediction, there is as yet no convincing signal of a contribution to CP 
violation from physics beyond the SM. Such a signal may eventually be provided 
by yet more precise measurements and joint analysis of CP asymmetries measured 
in different channels. Equally important is to look for CP violation in 
channels in which no (or negligibly small) effects violating CP are predicted 
by the SM. Observation of a non-zero effect in such processes would be an 
unambiguous signal of new physics contribution to CP violation.

{}From many possible channels\footnote{For example, also very small CP 
violation is predicted by the SM in the charm sector.}
we consider in this letter flavour changing decays of the neutral $B$ mesons
into lepton pairs, $B^0_{d,s}\rightarrow l^+l^-$. This channel has attracted 
a lot of attention since it is very sensitive to new physics which affects
the $b$-quark Yukawa couplings \cite{KASK,HAPOTO,BAKO,CHSL}. Approximate and 
full one-loop calculations in the supersymmetric extension of the SM with 
large $\tan\beta$ (the ratio of the vacuum expectation values of the two Higgs
doublets) \cite{BAKO,CHSL,ISRE} showed that in such a scenario one can expect
truly spectacular enhancement of the rates of the decays 
$B^0_{d,s}\rightarrow l^+l^-$.  

Moreover, new physics can also lead to observable CP violation in these decays,
which is not
predicted by the SM. For example, CP violation could manifest itself through
non-equal leptonic decay rates of the tagged $B^0(t)$ and $\bar B^0(t)$ states
($B^0(t)$ and $\bar B^0(t)$ are the states which at $t=0$ are tagged
as $B^0$ and $\bar B^0$, respectively). If polarization of final state
leptons can be determined, additional information on CP violation could be
provided by non-equal
$\Gamma(B^0(t)\rightarrow l^+_Ll^-_L)$ and
$\Gamma(\bar B^0(t)\rightarrow l^+_Rl^-_R)$ [or
$\Gamma(B^0(t)\rightarrow l^+_Rl^-_R)$ and
$\Gamma(\bar B^0(t)\rightarrow l^+_Ll^-_L)$] decay rates
\cite{Huang:2000tz,Dedes:2002er}.

Theoretically leptonic decays are particularly clean as the only
nonperturbative quantities, on which their rates depend, are the $B^0$
meson decay constants $F_{B_{d,s}}$. Moreover, $F_{B_{d,s}}$ cancel
out in suitably defined CP asymmetries. On the other hand it is not
clear whether the CP violation in these channels can be accessible
experimentally. First of all these decays have not yet been seen: the
best upper limit on the $B^0_d\rightarrow\mu^+\mu^-$ branching
fraction comes at present from the \babar\ experiment
\cite{Babar_Bdlim}:
\begin{eqnarray} 
Br(B^0_d\rightarrow\mu^+\mu^-)<8.3\times10^{-8}~,\label{eqn:Bdlim}
\end{eqnarray}
which improves on previous limits of BELLE ($1.6\times10^{-7}$ 
\cite{BELLE_Bdlim}) and  CDF ($1.5\times10^{-7}$ \cite{CDF_Bdlim}). 
The $B^0_s\rightarrow\mu^-\mu^+$ branching fraction is bounded by 
\cite{CDFD0} 
\begin{eqnarray} 
Br(B^0_s\rightarrow\mu^+\mu^-)<2.7\times10^{-7}\label{eqn:Bslim}
\end{eqnarray}
resulting from the combination of the CDF ($5.8\times10^{-7}$
\cite{CDF_Bdlim}) and D0 ($4.1\times10^{-7}$ \cite{D0_Bslim}) (all
limits are at 90\% CL).  The limits (\ref{eqn:Bdlim}) and
(\ref{eqn:Bslim}) are still about 2-3 orders of magnitude above the
corresponding predictions of the SM:
$Br(B^0_d\rightarrow\mu^+\mu^-)\approx1.3\times10^{-10}$ and
$Br(B^0_s\rightarrow\mu^+\mu^-)\approx3.6\times10^{-9}$
\cite{BURAS,Ball:2000ba,CHSL2}. The main
uncertainties of these predictions come from the decay constants
$F_{B_s}$ and $F_{B_d}$ which are known up to a precision of $\sim 15\%$
\cite{Bdec}. 
If the decays $B^0_{d,s}\rightarrow\mu^+\mu^-$ occur at the rates as
predicted by the SM, their detection will become possible only at the LHC. 
New physics (like supersymmetry) can increase significantly their rates to 
a level that they can be observed at \babar, BELLE or Tevatron in near 
future. However, as we will show, in this case the ratio of time 
integrated leptonic decay rates of the $B^0(t)$ and $\bar{B}^0(t)$ states 
to $\mu^+\mu^-$ is  unlikely to deviate appreciably from unity. Polarization 
measurement also seems very difficult in the case of the $\mu^+\mu^-$ channel
\cite{Huang:2002}. In the $\tau^+\tau^-$ channel the situation is quite 
different: large CP violating effects can be expected, and $\tau$ 
polarization measurement is possible.  So far this channel has not been much 
explored experimentally on account of difficulties with identifying $\tau$ 
leptons. As a result, practically no experimental limits 
on $Br(B^0_{d,s}\to\tau^+\tau^-)$ exist despite the 
fact that the corresponding rates are expected 
to be a factor $m_\tau^2/m_\mu^2\sim300$ larger than the ones for decays into
$\mu^+\mu^-$ final states.

The purpose of this paper is twofold. Firstly, we point out that in  
realistic scenarios of large $\tan\beta$ MSSM, where the rates of 
$B^0(\bar{B}^0)\rightarrow l^+l^-$ decays are significantly increased 
to a level measurable at the running \babar\ and BELLE experiments, the  
CP asymmetry in the $B^0_d(\bar{B}^0_d)\to\tau^+\tau^-$ channel can be 
quite large  and potentially measurable.
Secondly, we identify  realistic 
experimental observables in which the CP asymmetry can be detected.
In addition to the ratio of time integrated leptonic $B^0_d(t)$ and 
$\bar B^0_d(t)$ already mentioned, we consider two complementary observables: 
the $\pi^\pm$ energies from $\tau\to\pi\nu$ decays, and 
the acoplanarity angle between
the decay planes 
of the $\rho$ mesons which originate from $\tau\to\rho\nu$. The
former is sensitive to the longitudinal, while the latter to the
transverse polarizations of $\tau$'s  coming from $B^0_d$ and
$\bar B^0_d$ decays. We show how the existing {\tt TAUOLA} package 
\cite{Jadach:1990mz,Jezabek:1991qp,Jadach:1993hs} and its 
{\tt universal interface}  \cite{Pierzchala:2001gc,Was:2002gv,Golonka:2003xt} 
can be used to search for these decays {\it and} for the CP violation, 
demonstrating that the necessary tools for full Monte Carlo simulations  
are reliable and ready for use in 
the $B^0_d(\bar{B}^0_d)\rightarrow \tau^+\tau^-$ physics.

\section{General formulae}\label{sec:formulas}

The effective Lagrangian 
describing $B^0_d(\bar{B}^0_d)\rightarrow l^+l^-$ decays can succinctly 
be written as
\begin{eqnarray}
{\cal L}_{\rm eff} = B^0_{s,d}\bar\psi_l(b_{s,d}+a_{s,d}\gamma^5)\psi_l
         + \bar{B}^0_{s,d}\bar\psi_l(\bar b_{s,d}+\bar a_{s,d}\gamma^5)\psi_l~.
\label{eqn:leff}
\end{eqnarray}
To simplify the notation
the subscripts $d$ and $s$ referring to 
non-strange and strange $B^0$ mesons, unless explicitly written,  
will be omitted.
Hermiticity (CPT invariance) implies
\begin{eqnarray}
\bar b = b^\ast~,\phantom{aaaaaa}\bar a = -a^\ast~.\label{eqn:cpt_rel}
\end{eqnarray}
The amplitudes of $B^0$ decays into  two helicity eigenstates read
\begin{eqnarray}
{\cal A}_L\equiv\langle l^+_Ll^-_L|B^0_{s,d}\rangle
=M_B\left(a + b~\beta\right)~,\nonumber\\
{\cal A}_R\equiv\langle l^+_Rl^-_R|B^0_{s,d}\rangle
=M_B\left(a - b~\beta\right)~,\label{eqn:ALRdefs}
\end{eqnarray}
where $\beta=(1-4m_l^2/M^2_B)^{1/2}$. Similar formulae with $a$ and $b$ 
replaced by $\bar a$ and $\bar b$, respectively, give the amplitudes
$\bar{\cal A}_L$ and $\bar{\cal A}_R$ for the corresponding $\bar{B}^0$ decays.
Evidently, if both coefficients $a$ and $b$ are simultaneously nonzero,
$\Gamma(B^0\rightarrow l^+_Ll^-_L)\neq \Gamma(B^0\rightarrow l^+_Rl^-_R)$
but this is not yet a signal of CP violation. CP is violated, for
example,  if 
\begin{eqnarray}
\Gamma(B^0\rightarrow l^+_Ll^-_L)\neq \Gamma(\bar{B}^0\rightarrow l^+_Rl^-_R)~,
\label{eqn:cpt_viol1}\\
\Gamma(B^0\rightarrow l^+_Rl^-_R)\neq \Gamma(\bar{B}^0\rightarrow l^+_Ll^-_L)~,
\label{eqn:cpt_viol2}
\end{eqnarray}
because the initial and final states on both sides transform into each other 
under CP \cite{Huang:2000tz}. As there are no strong phases involved, this 
can occur only through the mixing of the $B^0$ and $\bar{B}^0$ mesons. In the 
standard formalism \cite{BLS} the state $B^0_{\rm phys}(t)$ 
($\bar{B}^0_{\rm phys}(t)$) which at $t=0$ is a pure $B^0$ ($\bar{B}^0$) 
evolves in time according to
\begin{eqnarray} 
|B^0_{\rm phys}(t)\rangle=g_+(t)|B^0\rangle+{q\over p}g_-(t)|\bar{B}^0\rangle~,
\nonumber\\
|\bar{B}^0_{\rm phys}(t)\rangle=g_+(t)|\bar{B}^0\rangle
+{p\over q}g_-(t)|B^0\rangle~.\nonumber
\end{eqnarray}
Neglecting the difference of the decay widths of the two $B^0$ mass 
eigenstates one finds 
\begin{eqnarray} 
g_+(t) = e^{-iM_Bt-{\Gamma\over2}t} ~\cos{\Delta M\over2} t~, 
\phantom{a}\nonumber\\
g_-(t) = e^{-iM_Bt-{\Gamma\over2}t} ~i ~\sin{\Delta M\over2} t~,
\end{eqnarray}
where $\Delta M\equiv M_{B_H^0}-M_{B_L^0}\ll M_B\equiv(M_{B_H^0}+M_{B_L^0})/2$.
The ratio $p/q$ (calculable in the SM or its extensions) is given by
\begin{eqnarray} 
{p\over q}=\sqrt{{H_{12}^\ast\over H_{12}}}
\approx{M_{12}^\ast\over |M_{12}|}\left(1-{1\over2}{\rm Im}
{\Gamma_{12}\over M_{12}}\right)~,
\end{eqnarray}
with $H_{12}\equiv M_{12}+{i\over2}\Gamma_{12}=
\langle B^0|{\cal H}_{\rm eff}|\bar{B}^0\rangle$, {\it etc.} \cite{BLS}.
The probability amplitude that the state, which initially 
was a $B^0$, decays at time $t$ into left-handed leptons is therefore given by
\begin{eqnarray} 
\langle l^+_Ll^-_L|B^0_{\rm phys}(t)\rangle=g_+(t)~{\cal A}_L
+ {q\over p}g_-(t)~\bar{\cal A}_L~,
\end{eqnarray}
and the rates of $B^0\rightarrow l^+_Ll^-_L$ and 
$\bar{B}^0\rightarrow l^+_Rl^-_R$ decays are proportional to
\begin{eqnarray} 
|\langle l^+_Ll^-_L|B^0_{\rm phys}(t)\rangle|^2=|g_+(t)|^2~|{\cal A}_L|^2
\left\{1+ \left|{q\over p}{g_-(t)\over g_+(t)}\right|^2~
\left|{\bar{\cal A}_L\over{\cal A}_L}\right|^2
+2{\rm Re}\left({q\over p}{g_-(t)\over g_+(t)}{\bar{\cal A}_L\over{\cal A}_L}
\right)\right\}~,\nonumber\\
|\langle l^+_Rl^-_R|\bar{B}^0_{\rm phys}(t)\rangle|^2
=|g_+(t)|^2~|\bar {\cal A}_R|^2
\left\{1+ \left|{p\over q}{g_-(t)\over g_+(t)}\right|^2~
\left|{{\cal A}_R\over\bar{\cal A}_R}\right|^2
+2{\rm Re}\left({p\over q}{g_-(t)\over g_+(t)}{{\cal A}_R\over\bar{\cal A}_R}
\right)\right\}~.\nonumber
\end{eqnarray}
The matrix elements for $B^0\rightarrow l^+_Rl^-_R$ and 
$\bar{B}^0\rightarrow l^+_Ll^-_L$ can be obtained from the ones above by
interchanging $L\leftrightarrow R$. 

Since $|{\cal A}_L|=|\bar {\cal A}_R|$, $|{\cal A}_R|=|\bar {\cal A}_L|$
and, as follows from (\ref{eqn:cpt_rel}) and (\ref{eqn:ALRdefs}),
\begin{eqnarray} 
{\bar{\cal A}_L\over{\cal A}_L}=
\left({{\cal A}_R\over\bar{\cal A}_R}\right)^\ast~,\label{eqn:basicrel}
\end{eqnarray}
one sees that the CP is violated if  either
\begin{eqnarray} 
\left|{q\over p}\right|\neq1\phantom{aaaaaa}{\rm or}
\phantom{aaaaaa}{\rm Im}\left(\lambda_L
\right)\neq {\rm Im}\left(\lambda_R^{-1}\right)~,\nonumber
\end{eqnarray}
where 
\begin{eqnarray} 
\lambda_L\equiv {q\over p}{\bar{\cal A}_L\over{\cal A}_L}~, 
\phantom{aaaa}{\rm and}\phantom{aaaa}
\lambda_R\equiv {q\over p}{\bar{\cal A}_R\over{\cal A}_R}~.
\label{eqn:lllrdefs}
\end{eqnarray}

The simplest quantities measuring the amount of CP violation
are the asymmetries constructed out of time integrated polarized 
decay rates
\begin{eqnarray} 
&&A_{\rm CP}^1(t_1,t_2)\equiv
{\int_{t_1}^{t_2} dt~\Gamma(B^0_{\rm phys}(t)\rightarrow l^+_Ll^-_L)
-\int_{t_1}^{t_2} dt~\Gamma(\bar{B}^0_{\rm phys}(t)\rightarrow l^+_Rl^-_R)\over
\int_{t_1}^{t_2} dt~\Gamma(B^0_{\rm phys}(t)\rightarrow l^+_Ll^-_L)
+\int_{t_1}^{t_2} dt~\Gamma(\bar{B}^0_{\rm phys}(t)\rightarrow l^+_Rl^-_R)}~,
\label{eqn:as1cp}
\\[1mm]
&&A_{\rm CP}^2(t_1,t_2)\equiv
{\int_{t_1}^{t_2} dt~\Gamma(B^0_{\rm phys}(t)\rightarrow l^+_Rl^-_R)
-\int_{t_1}^{t_2} dt~\Gamma(\bar{B}^0_{\rm phys}(t)\rightarrow l^+_Ll^-_L)\over
\int_{t_1}^{t_2} dt~\Gamma(B^0_{\rm phys}(t)\rightarrow l^+_Rl^-_R)
+\int_{t_1}^{t_2} dt~\Gamma(\bar{B}^0_{\rm phys}(t)\rightarrow l^+_Ll^-_L)}~.
\label{eqn:as2cp}
\end{eqnarray}
and the ratio of integrated unpolarized decay rates
\begin{eqnarray} 
R_l(t_1,t_2)\equiv
{\int_{t_1}^{t_2} dt~\Gamma(B^0_{\rm phys}(t)\rightarrow l^+l^-)\over
\int_{t_1}^{t_2} dt~\Gamma(\bar B^0_{\rm phys}(t)\rightarrow l^+l^-)}~.
\label{eqn:R_l}
\end{eqnarray}
The time interval $(t_1,t_2)$ can be chosen according to the 
experimental convenience. If the number of tagged events is not very 
large, or there is a large uncertainty in experimental determination 
of the decay time $t$, one can exploit the asymmetries
$A_{\rm CP}^1\equiv A_{\rm CP}^1(0,\infty)$ and 
$A_{\rm CP}^2\equiv A_{\rm CP}^2(0,\infty)$ and the ratio 
$R_l\equiv R_l(0,\infty)$
for which it is 
straightforward to obtain \cite{Huang:2000tz,CHSL1}
\begin{eqnarray} 
&&A_{\rm CP}^1=
{{1\over2}x^2\left(|\lambda_L^2|-|\lambda_R^{-2}|\right)
-x~{\rm Im}\left(\lambda_L-\lambda_R^{-1}\right)\over2+x^2
+{1\over2}x^2\left(|\lambda_L^2|+|\lambda_R^{-2}|\right)
-x~{\rm Im}\left(\lambda_L+\lambda_R^{-1}\right)}~,\label{eqn:as1cp_simpl}
\\[1mm]
&&A_{\rm CP}^2=
{{1\over2}x^2\left(|\lambda_R^2|-|\lambda_L^{-2}|\right)
-x~{\rm Im}\left(\lambda_R-\lambda_L^{-1}\right)\over2+x^2
+{1\over2}x^2\left(|\lambda_R^2|+|\lambda_L^{-2}|\right)
-x~{\rm Im}\left(\lambda_R+\lambda_L^{-1}\right)}~,\label{eqn:as2cp_simpl}
\end{eqnarray}
and
\begin{eqnarray} 
R_l={\left(|{\cal A}_L|^2+|{\cal A}_R|^2\right)
(1+{1\over2}x^2+{1\over2}x^2|{q\over p}|^2)
-x\left\{|{\cal A}_L|^2{\rm Im}(\lambda_L)
+|{\cal A}_R|^2{\rm Im}(\lambda_R)\right\}\over
\left(|{\cal A}_L|^2+|{\cal A}_R|^2\right)
(1+{1\over2}x^2+{1\over2}x^2|{p\over q}|^2)
-x \left\{|{\cal A}_L|^2{\rm Im}(\lambda_R^{-1})
+|{\cal A}_R|^2{\rm Im}(\lambda_L^{-1})\right\}}~,\label{eqn:Rl}
\end{eqnarray}
where
$ x\equiv \Delta M/\Gamma$. 
If $|q/p|=1$, the relation (\ref{eqn:basicrel}) implies 
$|\lambda_L|=|\lambda_R^{-1}|$. Moreover, $\lambda_L+\lambda_R^{-1}$ is 
then real, and the formulae (\ref{eqn:as1cp_simpl}), 
(\ref{eqn:as2cp_simpl}) and (\ref{eqn:Rl}) simplify to
\begin{eqnarray} 
&&A_{\rm CP}^1=
{-2~x~{\rm Im}~\lambda_L\over2+x^2+x^2~|\lambda_L|^2}~,\phantom{aaaaa}
A_{\rm CP}^2=
{-2~x~{\rm Im}\lambda_R\over2+x^2+x^2~|\lambda_R|^2}~,\nonumber
\\[1mm]
&& R_l={\left(|{\cal A}_L|^2+|{\cal A}_R|^2\right)(1+x^2)
-x\left\{|{\cal A}_L|^2{\rm Im}(\lambda_L)
+|{\cal A}_R|^2{\rm Im}(\lambda_R)\right\}\over
\left(|{\cal A}_L|^2+|{\cal A}_R|^2\right)(1+x^2)
+x \left\{|{\cal A}_L|^2{\rm Im}(\lambda_L)
+|{\cal A}_R|^2{\rm Im}(\lambda_R)\right\}}~.\label{eqn:Rlsimpl}
\end{eqnarray}
The asymmetries $A_{\rm CP}^1$, $A_{\rm CP}^2$, as functions of
$\lambda_{L,R}$, are bounded from above by \cite{Huang:2000tz}
\begin{eqnarray} 
\left|A_{\rm CP}^{1,2}\right|\leq{1\over\sqrt{2+x^2}}~.
\end{eqnarray}
Since $x_s>20.6$ for the $B^0_s$-$\bar{B}^0_s$ system, the CP asymmetries
in the leptonic $B^0_s(\bar{B}^0_s)$ decays can reach at best $\sim4.5$\%.
In contrast, for the $B^0_d$-$\bar{B}^0_d$ system, for which 
$x_d=0.771\pm0.012$ they can be as large as $\sim60$\%. This is fortunate,
since $B^0_d\bar{B}^0_d$ are copiously produced at \babar\ and BELLE in a 
relatively clean environment (compared to the $B^0$ production 
at hadron colliders). For this reason we will consider only the CP asymmetries
in the $B^0_d(\bar{B}^0_d)\rightarrow l^+l^-$ decays.

The quantities (\ref{eqn:as1cp}) and (\ref{eqn:as2cp}) 
depend on asymmetries of $B^0$ and $\bar B^0$ decays into longitudinally 
polarized leptons. In  the case of the $\tau^+\tau^-$ decay mode they   
are best identified by  measuring  the $\pi^\pm$ energy 
spectra from $\tau\rightarrow\pi\nu$ decays \cite{taupol}. 
The density 
matrix formalism outlined in Section \ref{sec:spindensity}  
allows to construct also observables 
sensitive to transverse polarization of the final state $\tau$'s 
\cite{Desch:2003rw}.  
These observables will prove  complementary since in some scenarios the 
signal of CP violation can clearly be visible in the latter observables while
hidden in the former (as $R_l$ can be expressed in terms of polarized
decay rates, $R_l$ 
     is then equal 1).

\section{Supersymmetry scenario}\label{sec:susy}

If the Cabibbo-Kobayashi-Maskawa (CKM) matrix is the
only source of {\it both} flavour and CP violation (as in the SM), then
$q/p\approx-V^\ast_{tb}V_{td(s)}/V_{tb}V^\ast_{td(s)}$ implying
$|q/p|\approx1$. Moreover, ${\cal A}_L,{\cal A}_R\propto
V^\ast_{tb}V_{td(s)}$, $\bar{\cal A}_L,\bar{\cal A}_R\propto
V_{tb}V^\ast_{td(s)}$, so that $\lambda_L$ and $\lambda_R$ are almost
real. The CP asymmetries in the $B^0\rightarrow l^+l^-$ decay are then
negligible.\footnote{In the SM they are estimated to be ${\cal
O}(10^{-2})$ \cite{Huang:2000tz} as a result of a small departure of
$|q/p|$ from $1$.} In models of new physics one can have ${\rm
Im}(\lambda_L)\neq{\rm Im}(\lambda_R^{-1})$ and/or $|q/p|\neq1$, but
if the predicted rates of these decays are still of the same order as
the SM predicts, the detection of the $\mu^+\mu^-$ decay mode
will become possible only at the LHC where most probably the
measurement of the muon polarization will not be feasible. 
Detecting CP violation in this mode could be possible then only
by measuring the ratio $R_\mu$. The
detection of the $B^0_d\rightarrow\tau^+\tau^-$ mode occurring at the
SM model level ($Br\sim5\times10^{-8}$) is possible at BELLE and
\babar\ \footnote{According to the SM prediction, among $5\times10^8$
events of $B^0_d\bar{B}^0_d$ pair production collected by these
experiments so far there should already be some 50 events of
$B^0_d(\bar{B}^0_d)$ decays into $\tau^+\tau^-$.}  but the number of
reconstructed events might be too small to detect any CP violation. 
On the other hand, at LHC where the number of
$B^0_{d,s}\rightarrow\tau^+\tau^-$ events will be larger, the
identification of the $\tau$ decay mode will probably be quite
difficult and the CP asymmetry hard to detect.

Much more promising situation can occur in the supersymmetric scenario with 
a large ratio of the vacuum expectation values of the two Higgs doublets, 
$v_u/v_d\equiv\tan\beta\sim40\div50$. The  coefficients $a$ and 
$b$ (and $\bar a$ and $\bar b$) in (\ref{eqn:leff}), and hence, the   
amplitudes of the 
$B^0_{d,s}\rightarrow l^+l^-$ decays can then receive important 
contributions from the Higgs penguin diagrams with $s$-channel  $H^0$ and 
$A^0$ Higgs boson  exchanges \cite{BAKO,CHSL,Dedes:2002er}.
If the mass scale of the Higgs particles $H^0$ and $A^0$ (for $v_u/v_d\gg1$ 
$M_H\approx M_A$) is not too high, of order $\simlt500$ GeV, 
the decay amplitudes 
can be dominated totally by these diagrams  
easily saturating  
the experimental limits (\ref{eqn:Bdlim}),
(\ref{eqn:Bslim}). This can happen even if the supersymmetric particle masses 
are quite large, say in the TeV range \cite{CHSL,BUCHROSL}.

As illustrative examples of new physics 
we consider here two different supersymmetric
scenarios: minimal (MFV) and non-minimal (NMFV) flavour violating, both
with large ratio of VEVs.

\noindent \underline{MFV Scenario:} ~ 
In this case no additional flavour violation in the sfermion mass 
matrices is assumed ({\it i.e.} the gluino-quark-squark vertices conserve 
flavour) but additional CP violation phases are  
introduced by complex parameters $\mu$ and $A_t$ (the higgsino mass term and 
left-right top squark mixing, respectively). The relevant effective flavour 
violating couplings of the two heavy neutral Higgs bosons to the down-type 
quarks\footnote{The 
effective flavour violating coupling of the lightest CP even Higgs $h^0$ 
to the down-type quarks is negligible.} 
can be written as 
\begin{eqnarray} 
{\cal L}^{H^0,A^0bd}_{\rm eff}=C~V_{tb}^\ast V_{td}~
\bar b\left[{m_b\over M_W}{A_t^\ast\over\mu}
(H^0+iA^0)\mathbf{P}_L +{m_d\over M_W}{A_t\over\mu^\ast}
(H^0-iA^0)\mathbf{P}_R\right] d\nonumber\\
+C~V_{td}^\ast V_{tb}~ \bar d\left[{m_d\over M_W}{A_t^\ast\over\mu}
(H^0+iA^0)\mathbf{P}_L +{m_b\over M_W}{A_t\over\mu^\ast}
(H^0-iA^0)\mathbf{P}_R\right] b ~,\nonumber
\end{eqnarray}
where $\mathbf{P}_{L,R}=(1\mp\gamma^5)/2$ are chiral projectors and the 
coefficient $C$ is given by
\begin{eqnarray} 
C= {g^3\over4}{m^2_t\over M_W^2}~\tan^2\beta~{H_2\over16\pi^2}\kappa\nonumber
\end{eqnarray}
with a dimensionless function of higgsino and stop masses $H_2$  of order 
${\cal O}(1)$. The factor $\kappa\propto\tan\beta $ 
summarises some refinements in the
calculation (resummation of $\tan\beta$ enhanced terms, for details see
\cite{ISRE,BUCHROSL}). It is the factor
$\tan^2\beta$ which makes these couplings so important. Combining these
couplings with the $H^0$ and $A^0$ couplings to $l^+l^-$ and using
\begin{eqnarray} 
\langle0|\bar b\mathbf{P}_Ld|B^0_d\rangle=-{i\over2}F_{B_d}{M^2_B\over m_b}~,
\phantom{aaaa}\langle0|\bar d\mathbf{P}_Lb|\bar{B}^0_d\rangle=
+{i\over2}F_{B_d}{M^2_B\over m_b}~, \nonumber\\
\langle0|\bar b\mathbf{P}_Rd|B^0_d\rangle=+{i\over2}F_{B_d}{M^2_B\over m_b}~,
\phantom{aaaa}\langle0|\bar d\mathbf{P}_Rb|\bar{B}^0_d\rangle=
-{i\over2}F_{B_d}{M^2_B\over m_b}~,\nonumber
\end{eqnarray}
one arrives at
\begin{eqnarray} 
a= -\bar a^\ast= C^\prime ~V_{tb}^\ast V_{td}~{M^2_B\over M_A^2}
\left({m_b\over M_W}{A_t^\ast\over\mu}
+{m_d\over M_W}{A_t\over\mu^\ast}\right) 
\approx C^\prime~V_{tb}^\ast V_{td}~{m_b\over M_W}
{M^2_B\over M_A^2}{A_t^\ast\over\mu}~,
\nonumber\\
b = \bar b^\ast= C^\prime ~V_{tb}^\ast V_{td}~{M^2_B\over M_H^2}
\left({m_b\over M_W}{A_t^\ast\over\mu}
-{m_d\over M_W}{A_t\over\mu^\ast}\right)
\approx C^\prime~V_{tb}^\ast V_{td}~{m_b\over M_W}
{M^2_B\over M_H^2}{A_t^\ast\over\mu}~,\nonumber
\end{eqnarray}
with 
\begin{eqnarray} 
C^\prime=-{g^4\over16}{m^2_t\over M_W^2}
{m_l\over M_W}{F_B\over m_b}~\tan^3\beta~
{H_2\over16\pi^2}\kappa\nonumber
\end{eqnarray}
In this case the amplitude of the $B^0_d$-$\bar{B}^0_d$ mixing is not
modified (in contrast to the one for $B^0_s$-$\bar{B}^0_s$ mixing)
\cite{BUCHROSL}, so that $|q/p|\approx 1$ holds true as in the SM. One
then finds 
\begin{eqnarray} 
\lambda_L\approx e^{-2i\delta_{\rm CP}}~{1-\beta\over1+\beta}~,
\phantom{aaaaaaa}
\lambda_R\approx e^{-2i\delta_{\rm CP}}~{1+\beta\over1-\beta}~,
\end{eqnarray}
where the effective CP violating phase is given as
$\delta_{\rm CP}={\rm arg}(\mu^\ast A_t^\ast)$. 
The time integrated asymmetries then read 
\begin{eqnarray} 
A^1_{\rm CP}=-0.09\times\sin(2\delta_{\rm CP})~,\phantom{aaaaa}
A^2_{\rm CP}=-0.35\times\sin(2\delta_{\rm CP})~.
\end{eqnarray}
Since the sparticles giving rise to substantial Higgs penguin contributions
to $a$ and $b$ ($\bar a$ and $\bar b$) can be quite heavy, even order 
${\cal O}(1)$ phase of $\mu A_t$ needs not produce unacceptable electric 
dipole moments.

\noindent \underline{NMFV Scenario:}
~ In this case  squark mass matrices violate flavour 
and the corrections generating the flavour changing couplings of 
the neutral Higgs bosons $H^0$ ad $A^0$ are dominated by gluino loops 
\cite{CHSL}. The relevant effective Lagrangian is then
\begin{eqnarray} 
{\cal L}^{H^0,A^0bd}_{\rm eff}=D~
\bar b\left[\alpha_b(H^0+iA^0)\mathbf{P}_L +\alpha_d^\ast
(H^0-iA^0)\mathbf{P}_R\right] d\nonumber\\
+D~ \bar d\left[\alpha_d
(H^0+iA^0)\mathbf{P}_L +\alpha_b^\ast
(H^0-iA^0)\mathbf{P}_R\right] b~, \label{eqn:bdHcoupl}
\end{eqnarray}
with
\begin{eqnarray} 
\alpha_b={m^\ast_{\tilde g}\over\mu}\left({m_b\over M_W}
\delta_{LL}^{bd}+{m_d\over M_W}\delta_{RR}^{bd}\right)~,
\phantom{aaaa}
\alpha_d={m^\ast_{\tilde g}\over\mu}\left({m_d\over M_W}
\delta_{LL}^{db}+{m_b\over M_W}\delta_{RR}^{db}\right)\nonumber
\end{eqnarray}
and
\begin{eqnarray} 
D={4\over3}g^2_sg\tan^2\beta{|\mu|^2\over M^2_{\tilde q}}
{H_3\over16\pi^2}~\kappa~, \nonumber
\end{eqnarray}
where $H_3$ is another dimensionless function of order ${\cal O}(1)$ of gluino
and sbottom masses and $\kappa$ again denotes dominant higher order 
contributions \cite{ISRE2}. The mass insertions $\delta_{LL}^{bd}$ and 
$\delta_{RR}^{bd}$ {\it etc.}  are the ratios of the diagonal entries of the 
down-type squark mass squared matrices to the average squark mass squared 
$M^2_{\tilde q}$. One then finds
\begin{eqnarray} 
a=-\bar a^\ast=D^\prime\left(\alpha_b+\alpha_d^\ast\right)
\approx D^\prime{m_b\over M_W}{M^2_B\over M^2_A}\left(
{m^\ast_{\tilde g}\over\mu}\delta_{LL}^{bd}
+{m_{\tilde g}\over\mu^\ast}\delta_{RR}^{db\ast}\right)~,\nonumber\\
b=\bar b^\ast=D^\prime\left(\alpha_b-\alpha_d^\ast\right)
\approx D^\prime{m_b\over M_W}{M^2_B\over M^2_H}\left(
{m^\ast_{\tilde g}\over\mu}\delta_{LL}^{bd}
-{m_{\tilde g}\over\mu^\ast}\delta_{RR}^{db\ast}\right)~,\phantom{a}\nonumber
\end{eqnarray}
with
\begin{eqnarray} 
D^\prime=-{1\over3}g^2_sg^2{m_l\over M_W}\tan^3
\beta{|\mu|^2\over M^2_{\tilde q}}
{H_3\over16\pi^2}~\kappa {F_B\over m_b}~. \nonumber
\end{eqnarray}
The insertions $\delta_{LL}^{bd}$ and $\delta_{RR}^{bd}$ are not very tightly 
constrained. Typically the bound of order
$|\delta_{LL(RR)}^{bd}|<0.2\times(M_{\tilde q}/$1~TeV) arising from 
$\Delta M_{B_d}$ is quoted \cite{GAGAMASI}. This estimate, however, 
does not take into account the contribution of the so-called double-penguin 
diagrams \cite{BUCHROSL0,CHRO,ISRE2,DE}, which can significantly affect the 
$B^0_d$-$\bar{B}^0_d$ mixing amplitude. This  contribution 
arises from the $H^0$ and $A^0$ 
exchanges between two effective (1-loop generated) 
vertices (\ref{eqn:bdHcoupl}) of which one vertex 
annihilates a right-chiral and the other  a left-chiral 
quark\footnote{Contributions 
of $H^0$ and $A^0$ exchanges between vertices 
annihilating same chirality quarks to the mixing amplitude  
are proportional to 
$1/M^2_H-1/M^2_A\approx0$ \cite{BUCHROSL0}.} and is therefore proportional to 
$\alpha_b\alpha_d^\ast\propto\delta_{LL}^{bd}\delta_{RR}^{db\ast}=
\delta_{LL}^{bd}\delta_{RR}^{bd}$   
(with the small d-quark mass neglected). 
Since this product is constrained much  stronger \cite{ISRE2},  
to avoid a potential conflict with 
the value of $\Delta M_{B_d}$ and the time dependent CP asymmetry 
$a_{J/\psi K_S}(t)$ measured in $B\rightarrow J/\psi K_S$ decay, we  
assume 
that either $\delta_{RR}^{bd}\gg\delta_{LL}^{bd}$ or 
$\delta_{RR}^{bd}\ll\delta_{LL}^{bd}$. This leads  to $a\approx 
\pm b$  and, as in 
the previous scenario, to $|q/p|\approx1$.  The asymmetries then read
\begin{eqnarray} 
A^1_{\rm CP}=-0.35\times\sin(2\delta_{\rm CP})~,\phantom{aaaaa}
A^2_{\rm CP}=-0.09\times\sin(2\delta_{\rm CP})~,\nonumber
\end{eqnarray}
where now 
\begin{eqnarray} 
\delta_{\rm CP}={\rm arg}(V_{tb}^\ast V_{td})
-{\rm arg}(m_{\tilde g}\mu\delta_{LL(RR)}^{bd})~. \label{eqn:phi_nonmin}
\end{eqnarray}
It is interesting to note here that the CP asymmetries can be nonzero 
even if the phase of the CKM matrix remains the only source of CP
violation ({\it i.e.} all supersymmetric parameters are real). Moreover,
since  $|{\rm arg}(V_{tb}^\ast V_{td})|$ is of order 1, the total
phase violating CP needs not be small.\\

In both scenarios, in which  the $B^0_{d,s}\rightarrow l^+l^-$
amplitudes are 
dominated by the exchange of $H^0$ and $A^0$ Higgs bosons whose effective 
flavour violating couplings to $bd$ (or $bs$) differ only by a 
factor $i$, one gets  
\begin{eqnarray} 
a\approx b\phantom{aaa} {\rm or}\phantom{aaa} a\approx-b~, \nonumber
\end{eqnarray}
(up to $\simlt15\%$).
{}For $a=b$ the factors $\lambda_L$ and $\lambda_R$ defined 
in Eq.~(\ref{eqn:lllrdefs}) simplify to:
\begin{eqnarray} 
\lambda_L=-{q\over p}~{a^\ast\over a}{1-\beta\over1+\beta}~,
\phantom{aaaaaa}
\lambda_R=-{q\over p}~{a^\ast\over a}{1+\beta\over1-\beta}~
\label{eqn:lambdas}
\end{eqnarray}
and all CP-sensitive quantities depend on one effective 
phase which can be taken as
\begin{eqnarray}
\delta_{\rm CP}=-{1\over2}{\rm arg}(\lambda_L)~. \label{eq:CPphase}
\end{eqnarray}

The immediate consequence of $a=b$, with $|q/p|$ close to 1, is that for the 
$\mu^+\mu^-$ final state the parameters $|\lambda_L|$ and $|\lambda_R|$ assume 
values $\sim4\times10^{-4}$ and $\sim2.5\times10^3$, respectively, since in 
this case $\beta=(1-4m_\mu^2/M^2_B)^{1/2}$ is almost 1. As a result, the 
predicted asymmetries are very small: $|A_{\rm CP}^1|\simlt2\times10^{-4}$, 
$|A_{\rm CP}^2|\simlt10^{-3}$.
The same is true if $a$ and $b$ are somewhat split. 
In contrast, for the $\tau^+\tau^-$ final states, for which
$\beta=(1-4m_\tau^2/M^2_B)^{1/2}$ differs 
substantially from $1$, we have $|\lambda_L|\sim0.15$, $|\lambda_R|\sim6.7$
and the maximal possible values of the asymmetries are 
\begin{eqnarray} 
\left|(A_{\rm CP}^1)^{\rm max}\right|=9\% \phantom{aaa} {\rm and}\phantom{aaa} 
\left|(A_{\rm CP}^2)^{\rm max}\right|=35\%~.
\end{eqnarray}

\begin{figure}[!ht]
\begin{center}  
\epsfig{file=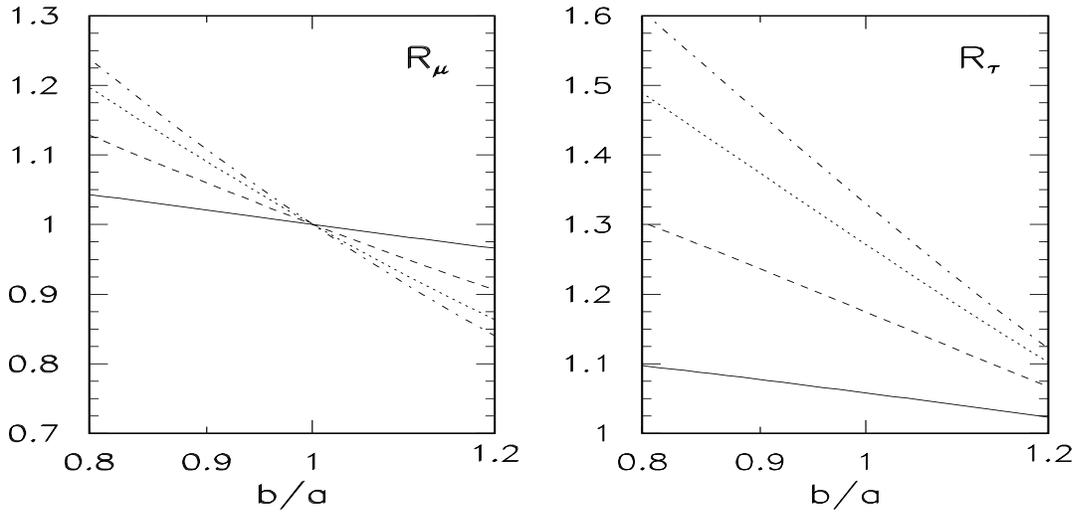,width=150mm,height=80mm}
\end{center} 
\caption  
{\it The ratios $R_\mu$ and $R_\tau$  as functions of 
$b/a$ for the phase $\delta_{\rm CP}=-{1\over2}{\rm arg}(\lambda_L)= 
0.1$ (solid line), $0.3$ (dashed), $0.5$ (dotted) and $0.75$ (dash-dotted). 
$R_l(-\delta_{\rm CP})=R^{-1}_l(\delta_{\rm CP})$.} 
\label{fig:Rl}  
\end{figure}

The comparison of magnitudes of possible CP violating effects in the ratio 
(\ref{eqn:Rl}) for  $\mu^+\mu^-$ and $\tau^+\tau^-$ decay modes is shown in
Fig.~\ref{fig:Rl}, where $R_\mu$ and $R_\tau$ are plotted as functions of 
$b/a$ for four different values of the phase $\delta_{\rm CP}$ (keeping 
arg$(a)=$arg$(b)$ and $|p/q|=1$). The plots show, that the ratios $R_l$ 
approach unity for $a\approx\pm b\beta$. (Vanishing of $R_l$ for 
$a\approx\pm b\beta$ follows also from the formula (\ref{eqn:Rlsimpl}) 
if one takes into account that  for $a\to b\pm\beta$
$\lambda_R\sim1/(a\mp b\beta)$ whereas $|{\cal A}_R|^2\sim|a\mp b\beta|^2$.)
Therefore, for $a\approx\pm b$ the deviation from unity of $R_\mu$ is tiny 
while for $R_\tau$ it can be quite substantial. 

The asymmetries $A^{1,2}_{\rm CP}$ (\ref{eqn:as1cp}), (\ref{eqn:as2cp}) and 
the ratio $R_l$ (\ref{eqn:R_l}) are the observables in which the signal of 
CP violation ({\it i.e.} a non-zero phase $\delta_{\rm CP}$) vanishes for 
$a\approx\pm b\beta$. As the spin density matrix formalism will show, 
transverse polarization of the final state leptons is free from this problem. 
However, observables sensitive to the transverse polarization can 
experimentally be constructed only for $\tau$'s which decay into 
hadrons. Thus, for $a\approx\pm b$ {\it only} the $\tau^+\tau^-$ channel 
provides an interesting opportunity to look for CP asymmetry in the leptonic
$B^0_d(\bar{B}^0_d)$ decays.

The coefficients $a$ and $b$ ($\bar a$ and $\bar b$) are constrained by the 
experimental limit Eq.~(\ref{eqn:Bdlim}), which in the case $a\approx 
\pm b$ gives
\begin{eqnarray}
|a|\approx|b|\simlt4.9\times10^{-9}~.\label{eqn:dirlim1}
\end{eqnarray}
In the  minimal flavour violation scenario, the parameters $a$
and $b$ are also constrained indirectly\footnote{ The limit imposed 
by the $B^0_s$-$\bar{B}^0_s$ mass difference
\cite{BUCHROSL0,BUCHROSL} can be avoided on account of a different 
$\tan\beta$ and $M_{A,H}$ dependence as compared to the  
the Higgs penguin diagram contributions 
to $B^0_d(\bar{B}^0_d)\rightarrow\ l^+l^-$ decays
\cite{CHSL2}.}  by the limit (\ref{eqn:Bslim}). Taking
$(m_\tau/m_\mu)(|V_{td}|/|V_{ts}|)\sim4$ this constraint is satisfied if
\begin{eqnarray}
|a|\approx|b|\simlt2\times10^{-9}~.\label{eqn:dirlim2}
\end{eqnarray}
In our simulations in Section \ref{sec:MC} 
we conservatively set $|a|=|b|\simlt10^{-9}$ and treat 
both scenarios simultaneously, as all what matters are the values of $|a|=|b|$ 
and the single CP violating phase $\delta_{\rm CP}$. In the MFV case 
$\delta_{\rm CP}={\rm arg}(\mu^* A_t^*)$ while in the NMFV case 
it is given by Eq.~(\ref{eqn:phi_nonmin}); 
in both cases the phase  can be of order 1.

Finally let us notice that the SM as well 
as supersymmetric box and $Z^0$ penguin contributions spoil the exact equality
$a=\pm b$. In addition, a finite difference of $A^0$ and $H^0$ masses also
splits these two coefficients. It is  therefore 
likely that $|a|$ and $|b|$ and the phases of $a$ and $b$ differ from
each other by some $10\div15$\% even for $|a|$ and $|b|$ saturating the bounds 
(\ref{eqn:Bdlim}), (\ref{eqn:Bslim}). 
Nevertheless, our simplified analysis  will demonstrate that in 
supersymmetry there are good reasons to expect substantial CP violation 
in $B_d^0(\bar{B}_d^0)\rightarrow\tau^+\tau^-$ decays. Therefore in the 
following sections we present tools which can be used 
to search for these decays and look for CP violation.

\section{Spin density matrix formalism}\label{sec:spindensity}

To see how the CP asymmetry in  
$B_d^0(\bar{B}_d^0)\rightarrow\tau^+\tau^-$ decays are reflected in
realistic observables we use the {\tt TAUOLA} $\tau$-lepton decay library
which allows us to simulate fully the effects of  $\tau$ polarization.
The exhaustive description of the method and  numerical algorithm is 
given in papers \cite{Jadach:1990mz} and \cite{Was:2002gv}. 
The input to the {\tt TAUOLA Universal Interface} \cite{Golonka:2003xt} 
is the spin density matrix of the $\tau^+\tau^-$
system resulting from the decay of a neutral particle. In this section
we collect the necessary formulae for this matrix for a
$\tau^+\tau^-$ pair originating from $B^0(\bar{B}^0)$.

The time dependent $B^0$-meson mixing can easily be 
dealt with by introducing time dependent effective factors $a_{\rm eff}$, 
$b_{\rm eff}$, $\bar a_{\rm eff}$ and $\bar b_{\rm eff}$ defined as:
\begin{eqnarray} 
&&a_{\rm eff}(t) = a ~g_+(t) + \bar a ~{q\over p}~g_-(t)
                 = a ~g_+(t) - a^\ast ~{q\over p}~g_-(t)~,\nonumber\\
&&b_{\rm eff}(t) = b ~g_+(t) + \bar b ~{q\over p}~g_-(t)
                 = b ~g_+(t) + b^\ast ~{q\over p}~g_-(t)~,\nonumber\\
&&\bar a_{\rm eff}(t) = \bar a ~g_+(t) + a ~{p\over q}~g_-(t)
                      =-a^\ast ~g_+(t) + a ~{p\over q}~g_-(t)~,\nonumber\\
&&\bar b_{\rm eff}(t) = \bar b ~g_+(t) + b ~{p\over q}~g_-(t)
                      = b^\ast ~g_+(t) + b ~{p\over q}~g_-(t)~,\nonumber
\end{eqnarray}
so that the instantaneous $B^0$ widths into left- or 
right-handed $\tau$'s read
\begin{eqnarray} 
\Gamma(B^0_{\rm phys}(t)\rightarrow \tau^+_L\tau^-_L)={M_B\over16\pi}\beta
\left|a_{\rm eff}(t) + \beta ~b_{\rm eff}(t)\right|^2,\\
\Gamma(B^0_{\rm phys}(t)\rightarrow \tau^+_R\tau^-_R)={M_B\over16\pi}\beta
\left|a_{\rm eff}(t) - \beta ~b_{\rm eff}(t)\right|^2,
\end{eqnarray}
and those for $\bar{B}^0$ are given by similar formulae with 
$a_{\rm eff}(t),~b_{\rm eff}(t)$ replaced by 
$\bar a_{\rm eff}(t),~\bar b_{\rm eff}(t)$. CP is violated because in general  
$\bar a_{\rm eff}(t)\neq-a_{\rm eff}^\ast(t)$, and
$\bar b_{\rm eff}(t)\neq b_{\rm eff}^\ast(t)$.  

The spin weight for the  $\tau^+\tau^-$ pair originating from $B^0$ decay  
at time $t$  is given by
\begin{eqnarray}
wt = {1\over4}{T(s_1,p_1,s_2,p_2)\over T(0,p_1,0,p_2)}~,\label{eq:wt}
\end{eqnarray}
where 
\begin{eqnarray}
T(s_1,p_1,s_2,p_2)={\rm Tr}\left[P(s_1)
\left(\not\! p_1 +m_l\right)(b_{\rm eff}+a_{\rm eff}\gamma^5)P(s_2)
\left(\not\! p_2 -m_l\right)(b^\ast_{\rm eff}-a^\ast_{\rm
  eff}\gamma^5)\right]~, 
\nonumber
\end{eqnarray}
with $P(s)\equiv{1\over2}(1+\gamma^5\!\! \not\! s)$, and $p_1$, $s_1$ ($p_2$, 
$s_2$)  are the momentum and spin four-vectors of the $\tau^-$ ($\tau^+$) 
lepton, respectively. The $\tau$-lepton spin four-vectors $s_a$ in the $B^0$ 
($\bar{B}^0$) rest frame are related to the spin three-vectors 
$\vec{\sigma}_a$ in the $\tau_a$-lepton rest frames as follows\footnote{In 
the rest frame of  $B^0/\bar B^0$ the $z$-axis is aligned with $\tau^-$ 
momentum,  exactly as in {\tt TAUOLA Universal Interface}. However, here as 
a particle number 1 the $\tau^-$ lepton is taken (not $\tau^+$), resulting  
in the transposition of the matrix $R_{\mu\nu}$ in the 
{\tt TAUOLA Universal Interface}.} 
\begin{eqnarray}
&& s^0_1= {M_B\over2m_l}\beta~\sigma^z_1~, 
\phantom{aaaa}
s^0_2=-{M_B\over2m_l}\beta~\sigma^z_2~,
\phantom{aa} \nonumber \\
&& s^z_a={M_B\over2m_l} \sigma^z_a~,
\phantom{aaaaai} 
s^{x,y}_a = \sigma^{x,y}_a~.
\end{eqnarray}
Combining (\ref{eq:wt}) with the  $\tau$-lepton decay matrix elements, 
the $\tau$-lepton rest frame spin vectors $\vec{\sigma}_1$ and
$\vec{\sigma}_2$ get  replaced by the polarimetric vectors $\vec{h}_1$ 
and $\vec{h}_2$ which are determined   
solely by the dynamic of the 
considered $\tau$ decay process. 

The spin weight for the complete event ($B\to\tau\tau\to$ decay
products)  can be written in  the form 
\begin{eqnarray}
WT={1\over4}\left(1+\sum_{i=x,y,z}\sum_{j=x,y,z}R_{ij}~h^i_1~h^j_2
+\sum_{i=x,y,z}R_{i0}~h^i_1+\sum_{j=x,y,z}R_{0j}~h^j_2
\right)~.\label{eqn:spwgh}
\end{eqnarray}
Expanding (\ref{eq:wt}) and comparing with (\ref{eqn:spwgh})  we find 
\begin{eqnarray}
\label{eq:matrixR}
&& R_{00}=+1, \qquad \qquad R_{x0}=R_{y0}=R_{0x}=R_{0y}=0, \nonumber \\
&& R_{zz}=-1, \qquad \qquad R_{xz}=R_{yz}=R_{zx}=R_{zy}=0,  \nonumber \\
&&R_{0z}=-R_{z0} = {2{\rm Re}(a_{\rm eff}b^\ast_{\rm eff})~\beta
\over|b_{\rm eff}|^2\beta^2+|a_{\rm eff}|^2},\\
&&R_{xy}=-R_{yx}=-{2{\rm Im}(a_{\rm eff}b^\ast_{\rm eff})~\beta
\over|b_{\rm eff}|^2\beta^2+|a_{\rm eff}|^2}, \phantom{aaaa}
R_{xx}=R_{yy}={|b_{\rm eff}|^2\beta^2-|a_{\rm eff}|^2
\over|b_{\rm eff}|^2\beta^2+|a_{\rm eff}|^2},\nonumber
\end{eqnarray}
Eq.~(\ref{eqn:spwgh}) with $R_{\mu\nu}$ replaced by 
$\bar R_{\mu\nu}$
computed as above but with $\bar a_{\rm eff}(t)$ and $\bar b_{\rm eff}(t)$
replacing $a_{\rm eff}(t)$ and $b_{\rm eff}(t)$, respectively, gives the spin 
weight for the events from $\bar{B}^0$ decays.  
CP violating effects in the $B^0\rightarrow\tau^+\tau^-$ and
$\bar B^0\rightarrow\tau^+\tau^-$ decays
are absent if $R_{0z}=\bar R_{z0}$,  
$R_{xy}=\bar R_{yx}$ and  
$R_{xx}=\bar R_{xx}$.

For simulations of the time integrated measurements, the time 
averaged matrix $\langle R_{\mu\nu}\rangle$ has to be used
\footnote{Simulations for time non-integrated measurements, with
  time-dependent decays  explicitly generated are also possible.}
\begin{eqnarray}
\langle R_{\mu\nu}\rangle\equiv
{\int dt ~\Gamma(B^0_{\rm phys}(t)\rightarrow \tau^+\tau^-) ~
R_{\mu\nu}(t)
\over\int dt ~\Gamma(B^0_{\rm phys}(t)\rightarrow \tau^+\tau^-)}
\nonumber
\end{eqnarray}
and $\langle\bar R_{\mu\nu}\rangle$ given by a similar formula.
The asymmetry (\ref{eqn:as1cp}) is then given by
\begin{eqnarray}
A_{\rm CP}^1={(1-\langle R_{z0}\rangle)\Gamma_{\rm int}
-(1+\langle\bar R_{z0}\rangle)\bar\Gamma_{\rm int}\over
(1-\langle R_{z0}\rangle)\Gamma_{\rm int} 
+(1+\langle\bar R_{z0}\rangle)\bar\Gamma_{\rm int}}~,
\label{eqn:asymbyR}
\end{eqnarray}
where
\begin{eqnarray}
\Gamma_{\rm int}=
\int dt ~\Gamma(B^0_{\rm phys}(t)\rightarrow
\tau^+\tau^-), 
\phantom{aaa}
\bar\Gamma_{\rm int}=
\int dt ~\Gamma(\bar B^0_{\rm phys}(t)\rightarrow \tau^+\tau^-).
\nonumber
\end{eqnarray}
$A_{\rm CP}^2(t_1,t_2)$ defined in (\ref{eqn:as2cp}) 
is given by (\ref{eqn:asymbyR}) reversing the signs in the brackets.  
It is also easy to check that for  $|q/p|=1$ and 
$a=\pm b$ one has
$\langle R_{z0}\rangle\Gamma_{\rm int}= -\langle\bar
R_{z0}\rangle\bar\Gamma_{\rm int}$
({\it i.e.} ${\rm Re}(a_{\rm eff}b^\ast_{\rm eff})
=-{\rm Re}(\bar a_{\rm eff}\bar b^\ast_{\rm eff})$).
Then the two factors: $\langle R_{z0}\rangle\Gamma_{\rm int}$ and 
$\langle\bar R_{z0}\rangle\bar\Gamma_{\rm int}$ cancel out in the numerator of
(\ref{eqn:asymbyR}) and of the similar formula for $A_{\rm CP}^2(t_1,t_2)$. 
As a result, nonzero asymmetries $A_{\rm CP}^{1,2}$ are possible only if 
$\Gamma_{\rm int}\neq\bar\Gamma_{\rm int}$, which in view of the equalities 
$|a_{\rm eff}(t)|=|\bar b_{\rm eff}(t)|$, 
$|b_{\rm eff}(t)|=|\bar a_{\rm eff(t)}|$ 
requires $\beta$ significantly different from $1$. This again 
confirms our observation made in Section \ref{sec:formulas} 
that for $a\approx\pm b$ 
the asymmetry in the $\mu^+\mu^-$ channel is suppressed.

Since in the limit $a=\pm b\beta$ the two integrated rates
$\Gamma_{\rm int}$ and 
$\bar\Gamma_{\rm int}$ 
are equal 
one gets $\langle R_{z0}\rangle=
-\langle\bar R_{z0}\rangle$, which means that  
even for $\delta_{\rm CP}\neq0$ there can be no CP violating effects in 
the observables sensitive to the longitudinal polarization of $\tau$'s
(nor in $R_\tau$). In 
contrast, it is easy to check by using the explicit analytical expressions,
that in this limit the elements $xx$ and $xy$ 
of these matrices {\it need not} satisfy 
$\langle R_{xy}\rangle=
-\langle\bar R_{xy}\rangle$ and  
$\langle R_{xx}\rangle=
\langle\bar R_{xx}\rangle$. 
Hence, the observables
sensitive to transverse $\tau$ polarization can reveal CP violation
even if the observables introduced in Section \ref{sec:formulas} fail
to signal it. 

\section{Results of the Monte Carlo simulations}\label{sec:MC}

As $B^0\rightarrow\tau^+\tau^-$ decays have not yet been seen it might
seem premature to study differential distributions in this channel,
including $\tau$ polarization. Nevertheless, our analysis can serve as
a good starting point for the future experimental work if indeed
accumulated experimental samples would turn to be large enough. 
By providing a Monte Carlo tool useful in
calculating {\it e.g.} detector responses, 
our study may also be considered
as benchmarks for the simulations to be used in setting the upper
limit for $B^0\to\tau^-\tau^+$ branching ratios.

There are some similarities between this study and the one 
aiming at  assessing measurability of the Higgs 
boson parity at future Linear Colliders \cite{Desch:2003rw}. 
The difference is that now the 
coefficients $a$ and $b$ (and also $\bar a$ and $\bar b$) in 
Eq.~(\ref{eqn:leff}) can be complex, while in \cite{Desch:2003rw} they were 
taken to be purely imaginary and real, respectively. As a result, in the 
formula Eq.~(\ref{eqn:spwgh}) terms linear in the polarimetric 
vector components also appear. Therefore we have extended the {\tt TAUOLA  
Universal Interface} to include such a possibility as well. The algorithm for 
simulating   $B^0$ decay into $\tau$ leptons is nearly identical to the 
one for  $H^0$ decay. 
Changes necessary for implementation in the {\tt Universal Interface}   
of the {\tt TAUOLA} 
Monte Carlo library  are 
limited to the replacement of Higgs identifier with the one for 
$B^0_d$ and $\bar{B}^0_d$ mesons, and the values of the
spin density matrix with the ones computed in the preceding section.

The formulae given in the previous section are general. However, motivated 
by the two supersymmetric scenarios discussed in Section \ref{sec:susy}, 
in our numerical 
study we will show first the results obtained in the limit $a=\pm b$. 
Then we will discuss effects of a small departure from this 
relation.

We will limit ourselves to two  observables, which are known to 
provide valuable and complementary 
information on the spin state of decaying $\tau$
lepton pairs. 

\noindent \underline{$\pi^\pm$ energies:}~ As the first observable 
we take the $\pi^+$ and $\pi^-$ energy spectra in the decay channels  
$\tau^+\to\pi^+\bar\nu_\tau$ (or $\tau^-\to\pi^-\nu_\tau$). Since they  
reflect the longitudinal polarization of the individual $\tau^\pm$ leptons, 
the spectra are sensitive to 
$R_{z0}$ and $\bar R_{0z}$ as can be inferred from the expression 
(\ref{eqn:spwgh}), {\it i.e.} they are sensitive to 
${\rm Re}(a_{\rm eff}b^\ast_{\rm eff})$ as follows from (\ref{eq:matrixR}). 
The CP violation is reflected in the difference between  
the energy spectrum of $\pi^-(\pi^+)$ originating from $B^0(\bar{B}^0)$ and 
the energy spectrum of $\pi^+(\pi^-)$ originating from $B^0(\bar{B}^0)$.
This observable 
was exploited before, for example in the $Z^0/\gamma^\ast\to\tau^+\tau^-$  
study (for a review of the method and its extensions, see {\it e.g.} 
Ref.~\cite{Heister:2001uh}). We will show energy spectra 
in the rest frame of the $B^0(\bar{B}^0)$ meson assuming that the 
reconstruction of the event kinematics in the BELLE and \babar\ experiments is 
sufficiently good for that purpose, that is, that the momenta of $\tau$ decay 
products in the rest frame of $B^0(\bar{B}^0)$ can be reconstructed with the
precision of a fraction of the $\tau$ mass.

\underline{Acoplanarity angle $\varphi^\ast$:}~ 
As the second observable we use the acoplanarity angle $\varphi^\ast$ 
between two planes spanned by the momenta of decay products of 
$\rho^\pm\to\pi^\pm\pi^0$ coming from decays of both $\tau$ leptons into 
$\rho\nu_\tau$ \cite{Desch:2003rw}. This quantity is sensitive to  
correlations between transverse components of $\tau$-lepton spins 
({\it i.e.} to the elements $R_{xx}$ and $R_{xy}$ which in turn 
probe ${\rm Im}(a_{\rm eff}b^\ast_{\rm eff})$, as can be seen from 
(\ref{eq:matrixR})). 
For the definition of the acoplanarity the orientation of decay planes 
and pion momenta has to be properly taken into account. 
The acoplanarity angle $\varphi^\ast$ is defined with the help of two 
vectors ${\bf n}_\pm$ normal to the planes determined by the momenta
of pions which originate from $\rho^\pm$ decays: 
${\bf n}_\pm={\bf p}_{\pi^\pm}\times{\bf p}_{\pi^0}$. If  
$\cos\xi ={{\bf n}_+ \cdot{\bf n}_-\over|{\bf n}_+||{\bf n}_-|}$ then 
\begin{eqnarray}
\varphi^\ast=\left\{ \begin{tabular}{cl}
                  $\xi$ &  ~~~~for~~ sgn$({\bf p}_{\pi^-} \cdot
                  {\bf n}_+)<0$ \\
                  $2\pi-\xi$ & ~~~~for~~ sgn$({\bf p}_{\pi^-} \cdot
                  {\bf n}_+)>0$
                  \end{tabular} \right.
\end{eqnarray}
making  the full range of the variable $0 <\varphi^\ast<2\pi$ of physical 
interest. Note that under CP, $\varphi^\ast \to 2\pi-\varphi^\ast$ since the 
condition sgn$({\bf p}_{\pi^-}\cdot{\bf n}_+)$ is always evaluated from
the orientation of $\pi^-$ momentum with respect to the vector ${\bf n}_+$. 
In addition, we also have to sort events depending whether    
$y_1 y_2>0$ or $y_1 y_2<0$, where   
\begin{equation}  
y_1={E_{\pi^+}-E_{\pi^0}\over E_{\pi^+}+E_{\pi^0}}~,\phantom{aaaaaaa}
y_2={E_{\pi^-}-E_{\pi^0}\over E_{\pi^-}+E_{\pi^0}}~,
\label{E-zone}   
\end{equation}
since otherwise the spin correlations are washed out, as explained in 
Ref.~\cite{Bower:2002zx}. The best would be to use in (\ref{E-zone}) 
the energies of $\pi^\pm$ and $\pi^0$'s in the rest frames of the 
corresponding $\tau^\pm$ leptons, but they are not directly measurable.
Since in the $B^0(\bar B^0)$ rest frame the $\tau$ leptons are only 
mildly relativistic, the difference of pion energies in this frame and 
respective rest frames of $\tau^\pm$ should not be very important. The 
acoplanarity distribution is then evaluated in the rest frame of the 
$\rho^+\rho^-$ pair, but with 
the energies of $\pi^\pm$ and $\pi^0$'s in (\ref{E-zone}) taken 
in the rest frame of the $B^0$($\bar B^0$). The CP violation is reflected 
in the difference between the distributions of the acoplanarity angle 
$\varphi^\ast$ measured in $B^0$ decays and  the angle $2\pi-\varphi^\ast$ 
measured in $\bar B^0$ decays for the same signs of $y_1y_2$.

\begin{figure}[!ht]
\begin{center}  
\epsfig{file=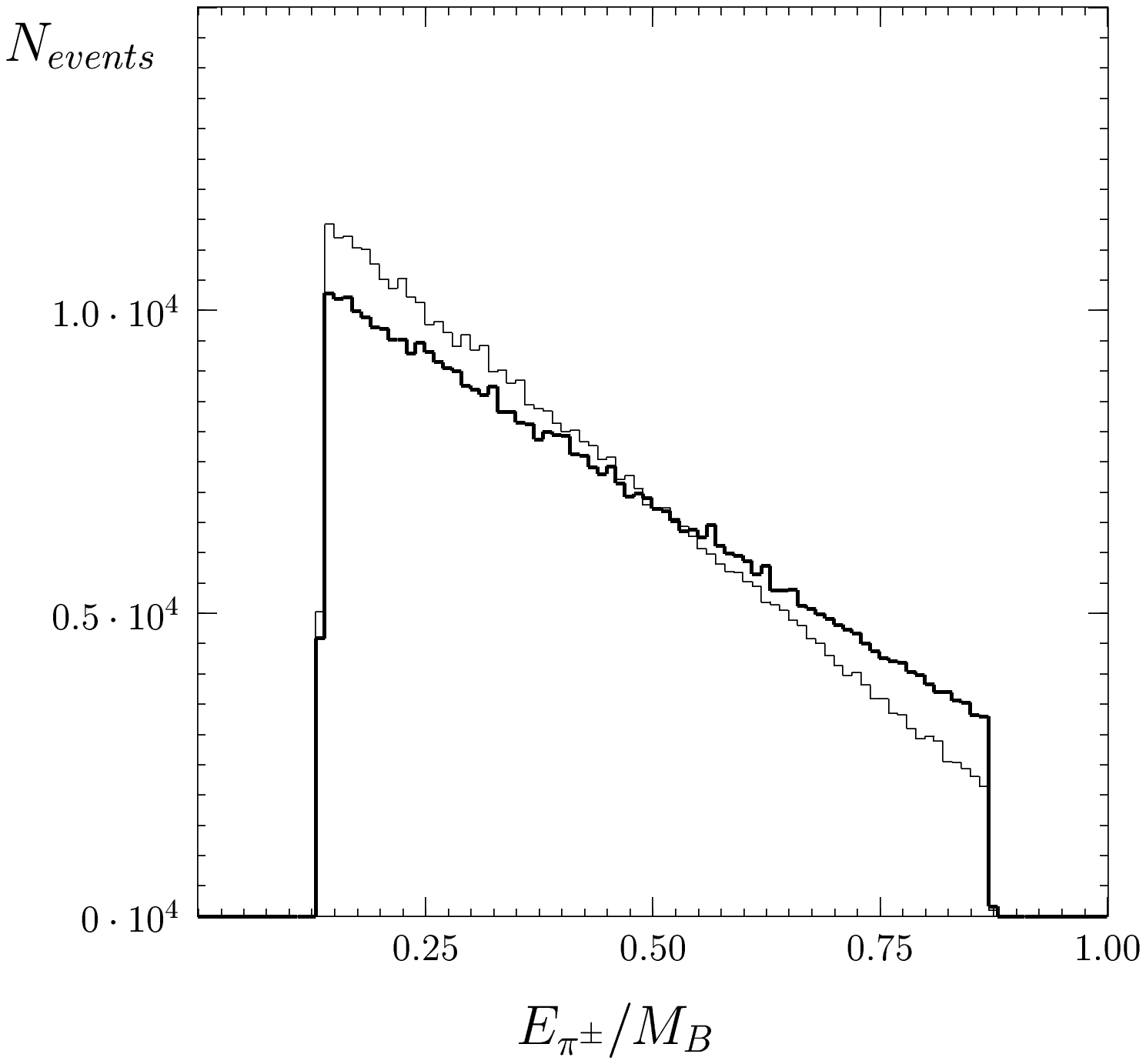,width=70mm,height=60mm}
\epsfig{file=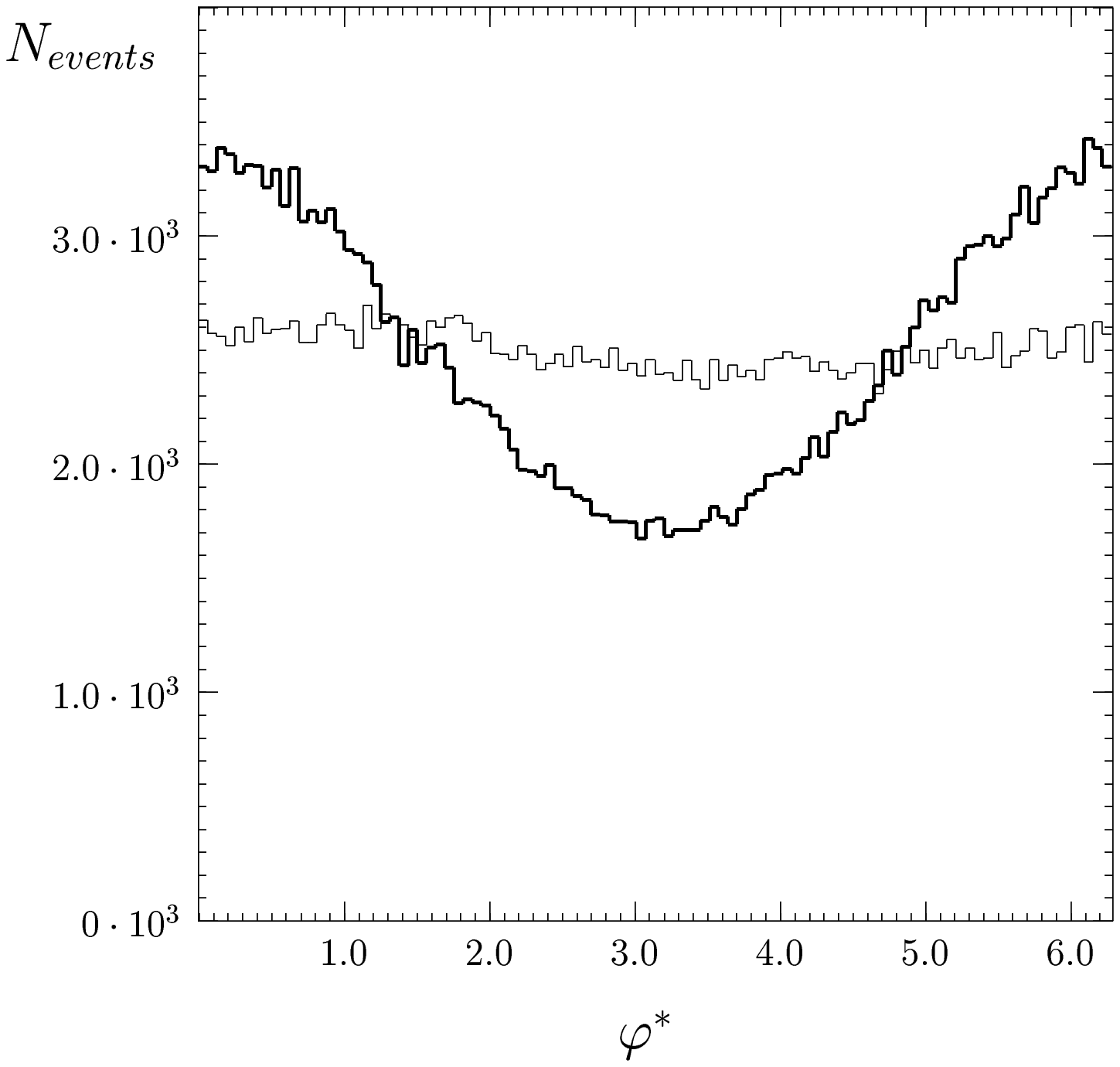,width=70mm,height=60mm}
\epsfig{file=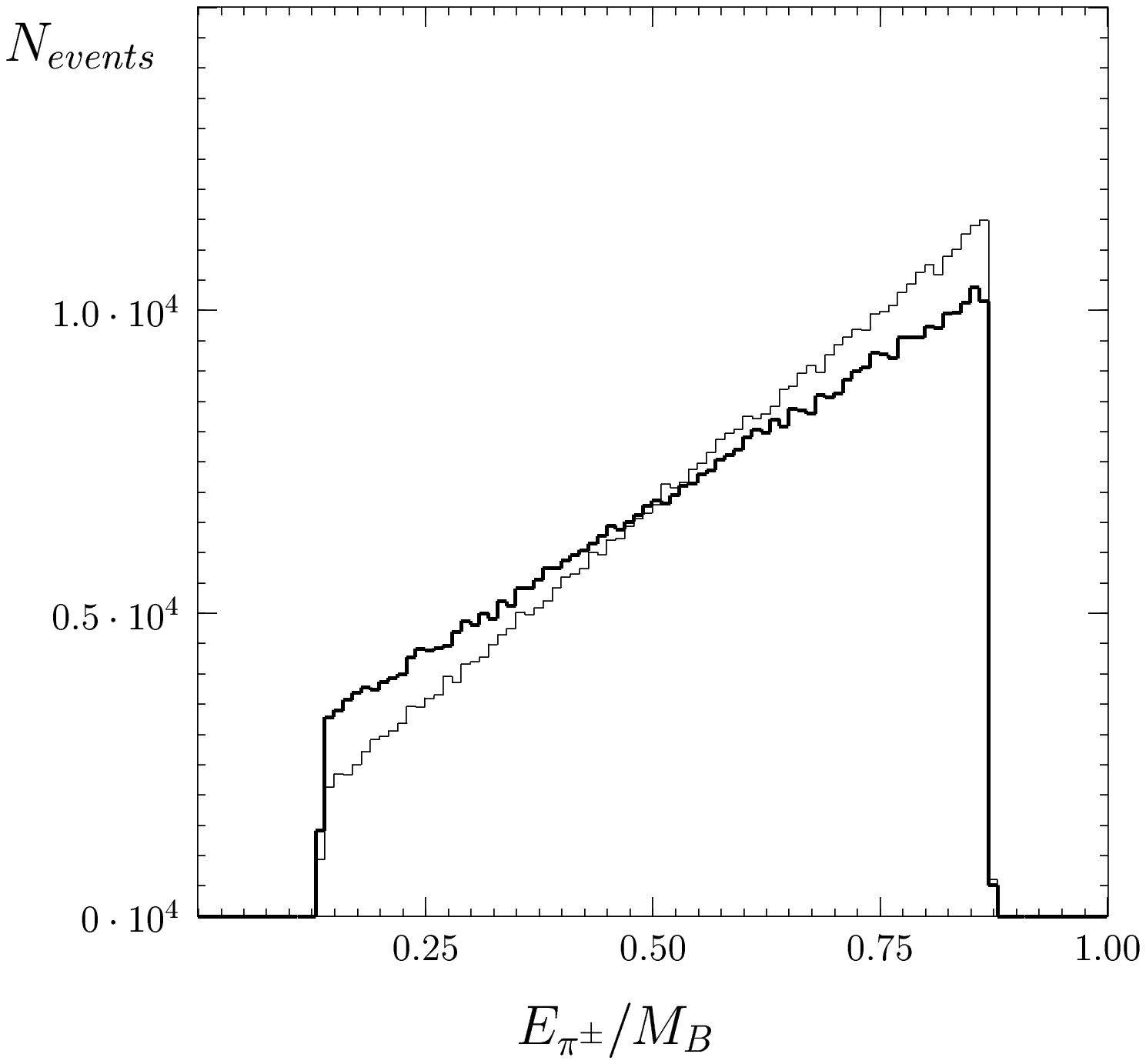,width=70mm,height=60mm}
\epsfig{file=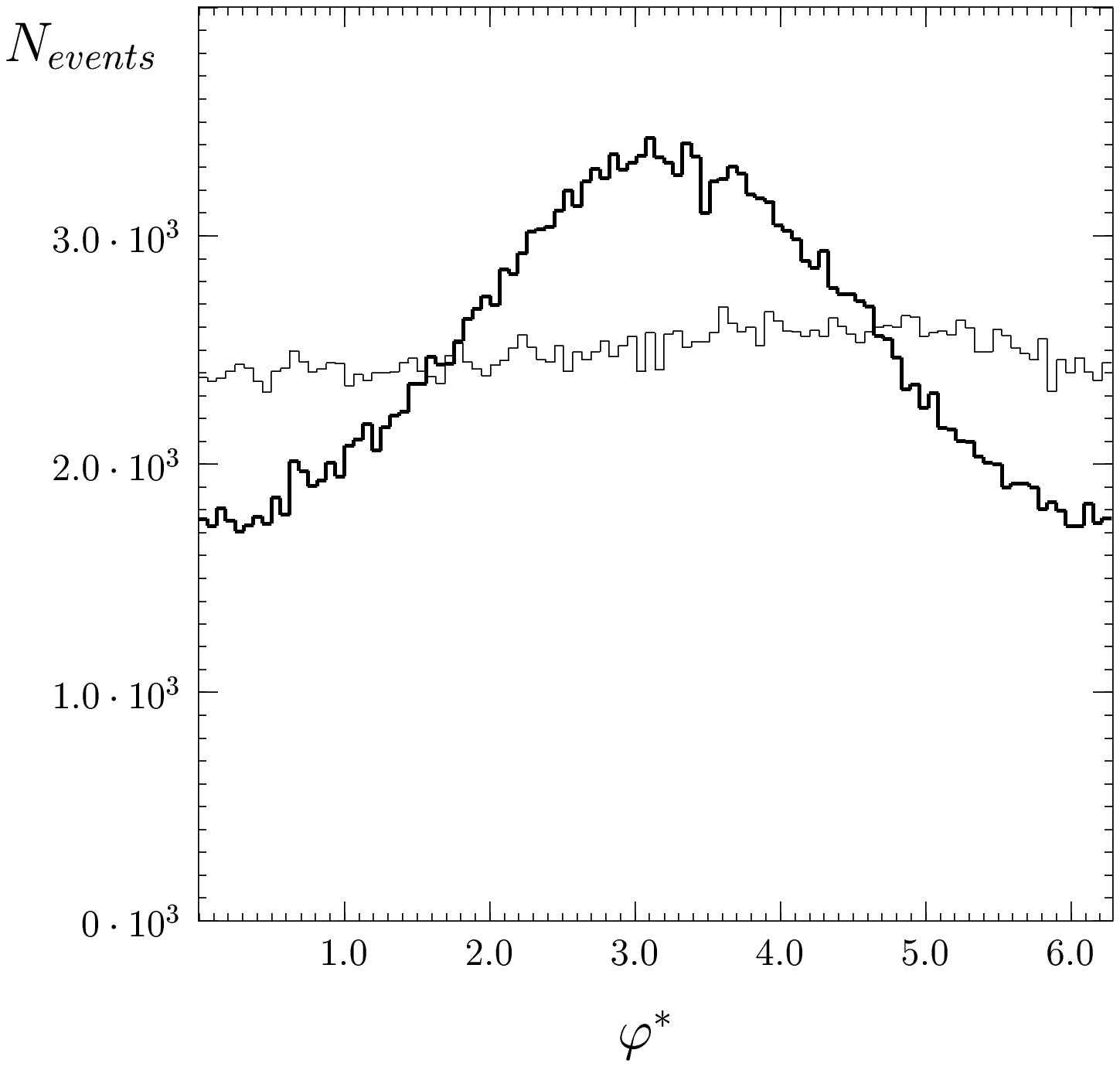,width=70mm,height=60mm}
\end{center} 
\caption  
{\it Results for the CP violating phase $\delta_{\rm CP}=0.7$ and $a=b$. 
Left panels:
Single $\pi^\pm$ energy spectra in $B^0(\bar B^0)\to\tau^+\tau^-$,
$\tau^\pm\to\pi^\pm\nu_\tau(\bar\nu_\tau)$. In the upper (lower) panel the 
thick line corresponds to the energy spectrum of $\pi^-$ (of $\pi^+$) from 
$B^0$ decays and the thin line to the energy spectrum of $\pi^+$ (of $\pi^-$) 
{}from $\bar{B}^0$. Spectra are plotted in the rest frame of $B^0(\bar B^0)$.
Right panels: acoplanarity distributions of the $\rho^+\rho^-$ decay products
in $B^0(\bar B^0)\to\tau^+\tau^-$, 
$\tau^\pm\to\rho^\pm\nu_{\tau}(\bar{\nu}_{\tau})$, 
$\rho^\pm\to\pi^\pm\pi^0$. The thick  lines correspond to the 
acoplanarity angle $\varphi^\ast$ measured in $B^0$ decays and the thin ones 
are for the angle $2\pi-\varphi^\ast$ measured in $\bar{B}^0$ decays. 
The acoplanarity angles are defined in the rest frame of the $\rho^+\rho^-$ 
pair. Events in the upper (lower) panel have $y_1 y_2 > 0$ ($y_1 y_2< 0$). } 
\label{fig:delta07}  
\end{figure}  

Fig.~\ref{fig:delta07} shows the pion energy spectra and the acoplanarity 
distributions assuming $|q/p|=1$, $a=b=10^{-9}$ and the CP violating phase 
$\delta_{\rm CP}=0.7$. For all plots the same number of $5\times10^5$ 
$\tau^+\tau^-$  events from $B_d^0$ and $\bar B_d^0$ decays has been 
generated with {\tt TAUOLA}, although for the parameters chosen the ratio 
$R_\tau= 1.32$, see Fig.~\ref{fig:Rl}. In the upper left panel the energy 
spectra of $\pi^-$ from $B^0$ decays (thick line with the slope proportional 
to $\langle R_{z0}\rangle$ \cite{Eberhard:1989ve}) 
and of $\pi^+$ from 
$\bar B^0$ (thin line; slope $\propto\langle\bar R_{0z}\rangle$) are shown, 
while in the lower left panel shown are the spectra of $\pi^+$ from $B^0$ 
decays (thick line; slope $\propto\langle R_{0z}\rangle$) and of $\pi^-$ from 
$\bar{B}^0$ (thin line; slope $\propto\langle\bar R_{z0}\rangle$).  
The harder $\pi^-$ energy spectrum from $B_d^0$ decays than $\pi^+$ from 
$\bar B_d^0$ ({\it i.e.} larger slope of the thick line) in the upper left 
panel indicates that 
$Br(B_d^0\to \tau^+_R\tau^-_R)>Br(\bar{B}_d^0\to \tau^+_L\tau^-_L)$, 
which is a clear signal of CP violation. In the acoplanarity plots (right 
panels) thick lines correspond to the distribution of $\varphi^\ast$ measured 
in $B^0$ decays, and the thin lines to the distribution of  
$2\pi-\varphi^\ast$ measured in $\bar{B}^0$ decays; in the upper right panel  
$y_1y_2>0$, and  $y_1y_2<0$ in the lower right one. The shapes of the thick 
and thin lines are described by the formulae 
\begin{eqnarray}
&& N_B(\varphi^\ast)={\rm const}
-{\rm sgn}(y_1y_2)A_R\cos(\varphi^\ast-\delta_R)\phantom{aaaaaaaa.}
{\rm for} \phantom{aa}B^0\to\tau^+\tau^-\nonumber\\
&& N_{\bar B}(\varphi^\ast)={\rm const}
-{\rm sgn}(y_1y_2)A_{\bar R}\cos(2\pi-\varphi^\ast+\delta_{\bar R})
\phantom{aaaa}
{\rm for} \phantom{aa}\bar B^0\to\tau^+\tau^-\nonumber
\end{eqnarray}
where $A_R=(R_{xx}^2+R_{xy}^2)^{1/2}$, $\sin\delta_R=R_{xy}/A_R$,
$\cos\delta_R=R_{xx}/A_R$ and $A_{\bar R}$ and $\delta_{\bar R}$ are given
by analogous formulae with $R_{ij}$ replaced by the $\bar R_{ij}$.
Different shapes of thick and thin lines seen in the right panels of
Fig.~\ref{fig:delta07} again indicate CP violation. In both energy and 
acoplanarity plots the CP violation is clearly seen and should be measurable 
even for small statistics. 

Note also that if upper and lower plots are combined ({\it i.e.} no 
sorting according to the pion charge or sign of $y_1y_2$ is made), 
all CP asymmetries are lost. 
Since the lower plots are simple reflections of the upper ones, in the 
following  only plots corresponding to the upper panels of 
Fig.~\ref{fig:delta07} are shown.
\begin{figure}[!ht]
\begin{center}  
\epsfig{file=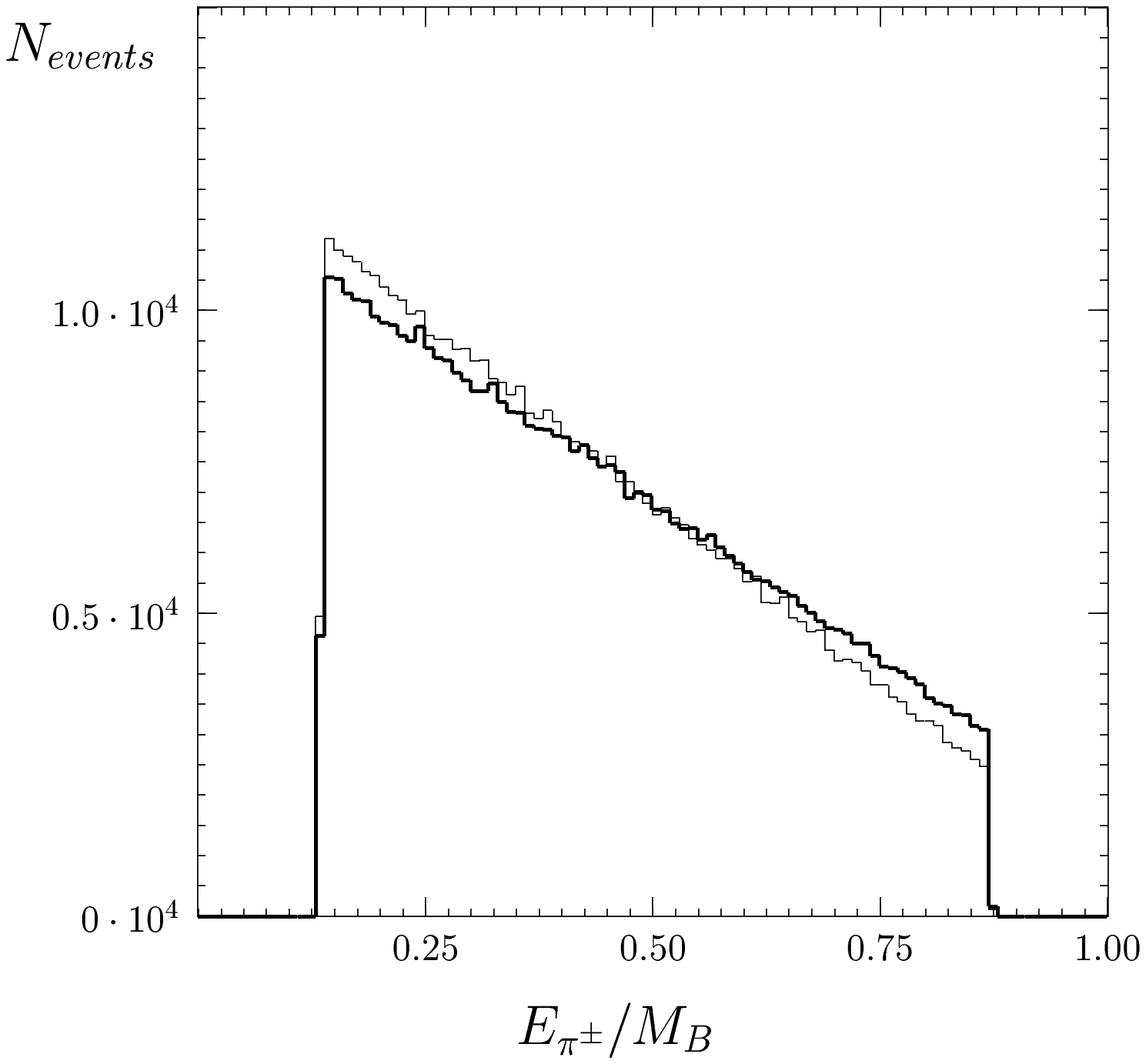,width=70mm,height=60mm}
\epsfig{file=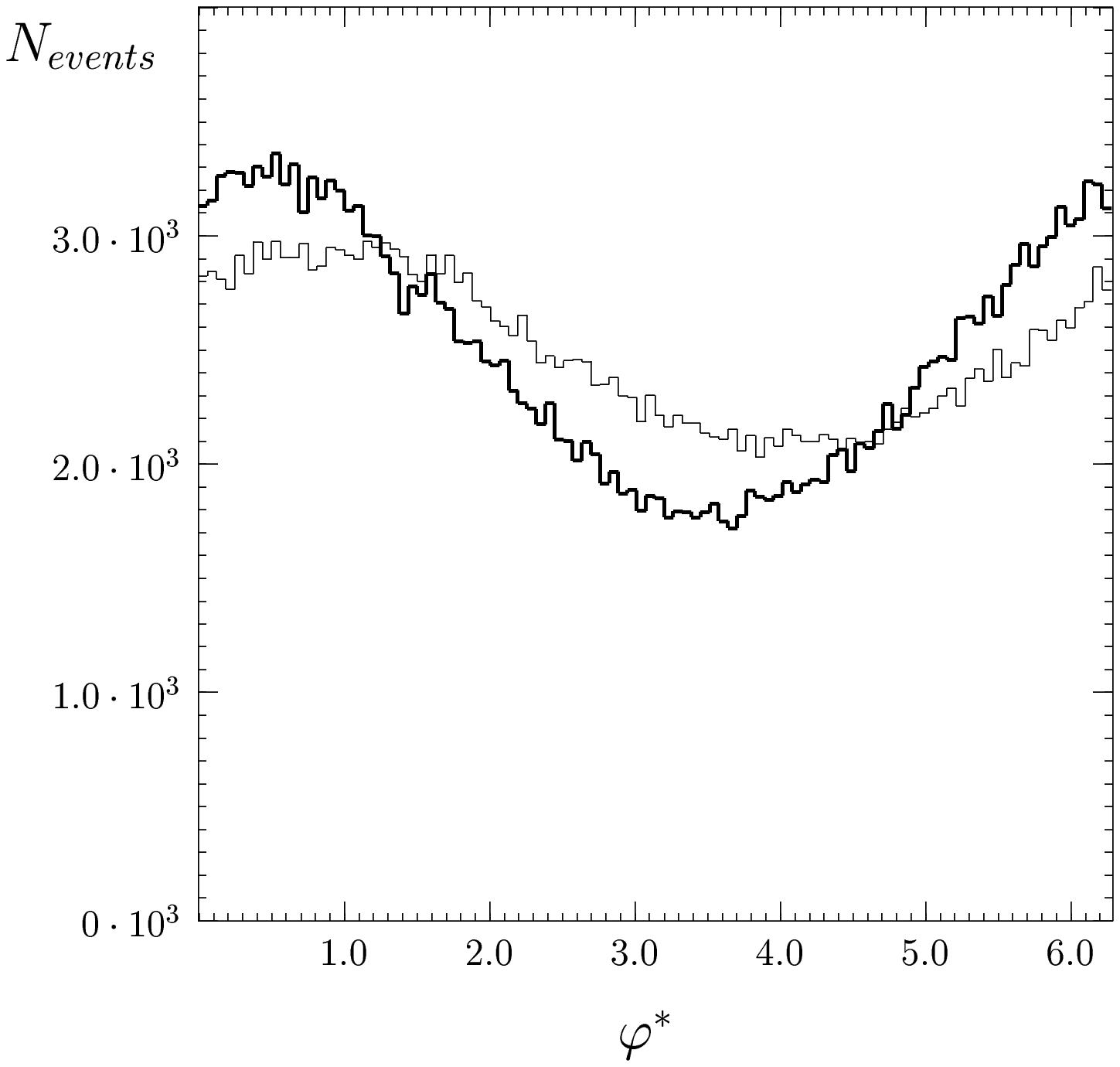,width=70mm,height=60mm}
\end{center} 
\caption 
{\it As in the upper panels of figure \ref{fig:delta07} but for
$\delta_{\rm CP}=0.3$ and $a=b$.}  
\label{fig:delta03}  
\end{figure}  

Fig.~\ref{fig:delta03} shows the corresponding distributions for the CP
violating phase $\delta_{\rm CP}=0.3$ keeping still $a=b$ and $|q/p|=1$. 
It is apparent that with decreasing $|\delta_{\rm CP}|$ the signal of CP
violation deteriorates (especially in the pion spectra)
and the possibility of
distinguishing  pion spectra and
acoplanarity distributions from $B^0$ and $\bar{B}^0$, 
and hence the CP violation, would require increasingly
large statistics which may not be attainable at BELLE and \babar\
without major upgrades.

\begin{figure}[!ht]
\begin{center}  
\epsfig{file=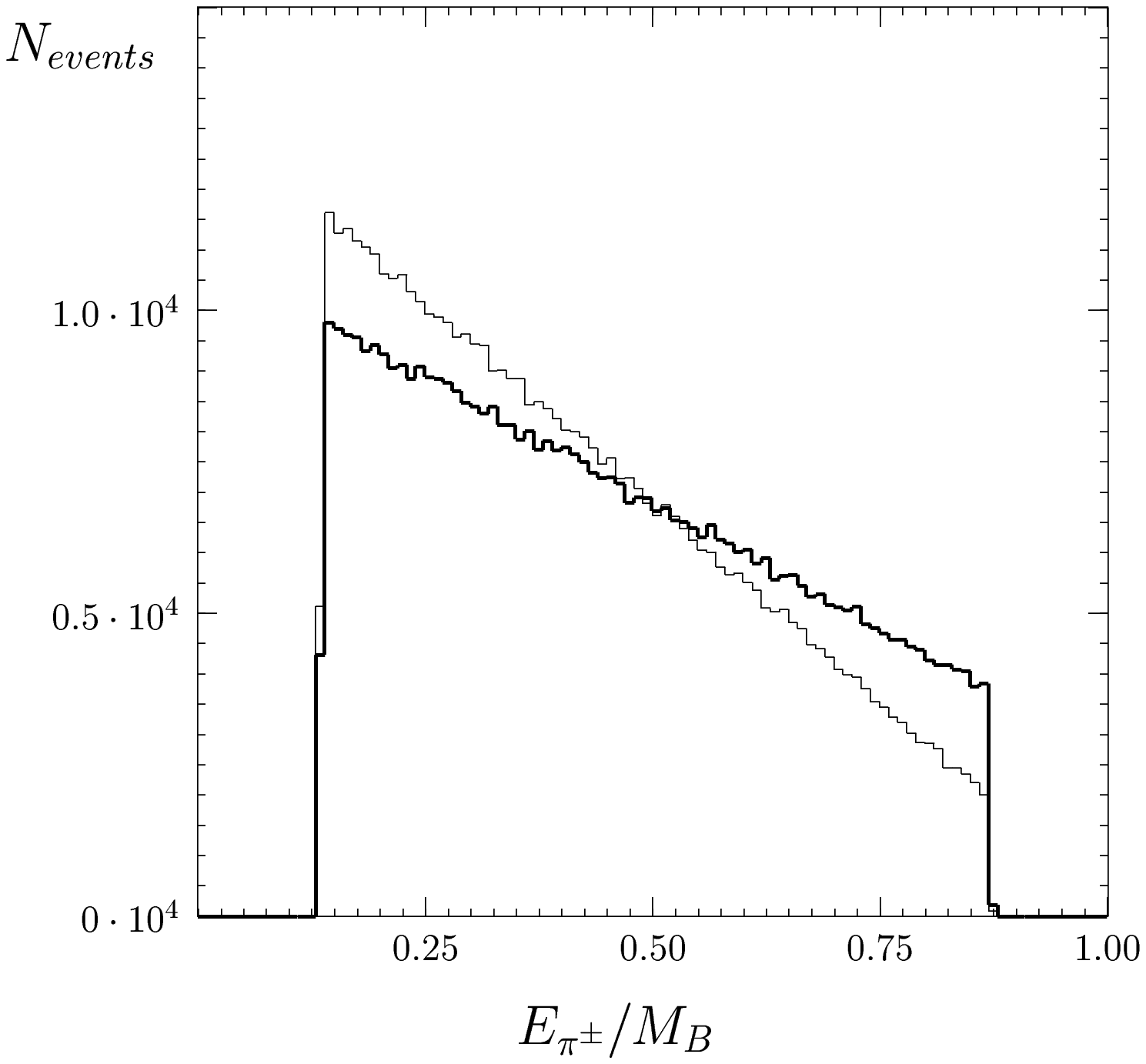,width=70mm,height=60mm}
\epsfig{file=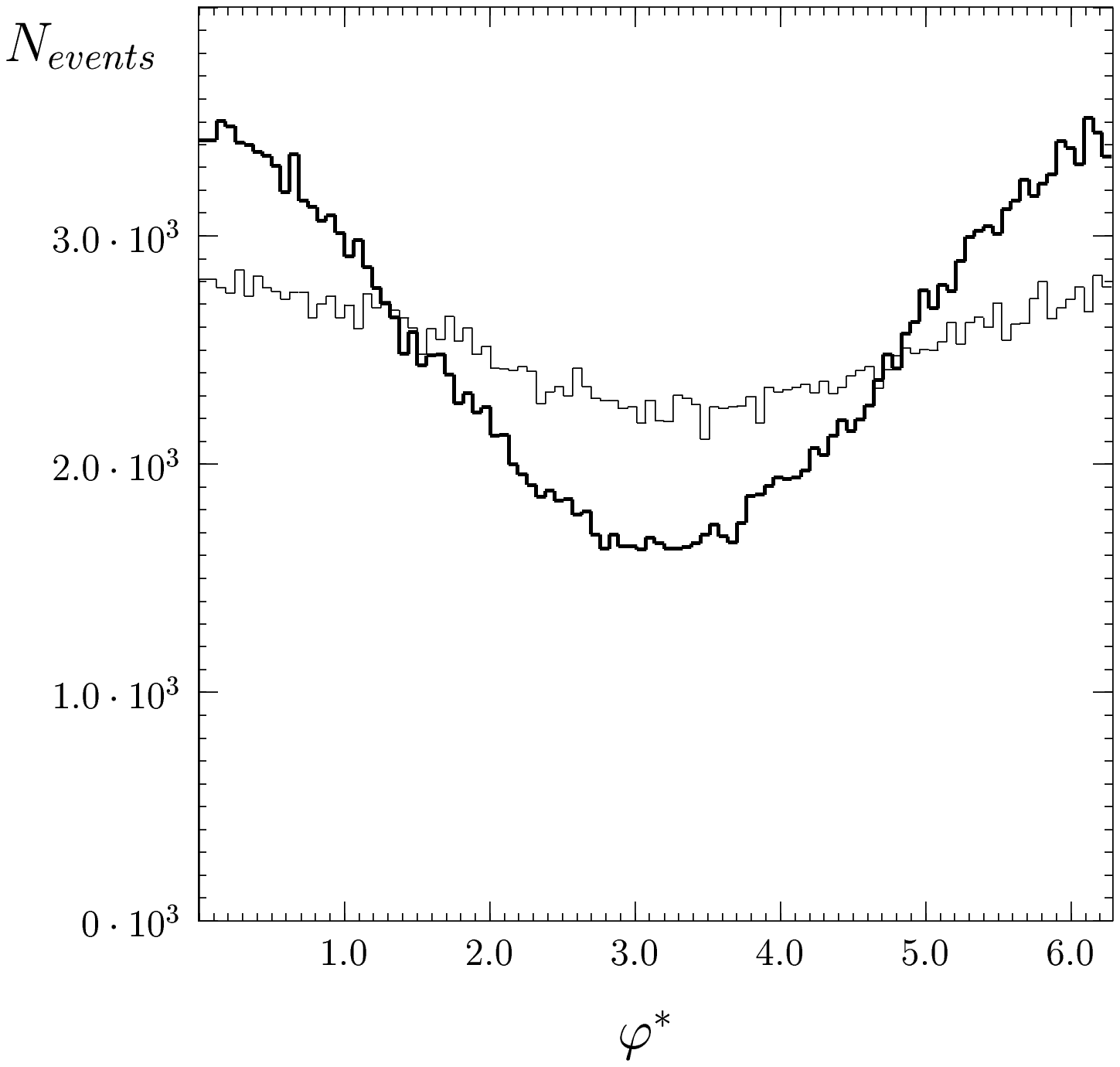,width=70mm,height=60mm}
\end{center} 
\caption 
{\it As in the upper panels of 
figure \ref{fig:delta07} but for $\delta_{\rm CP}=0.7$ and
  $b=0.8\, a$.}  
\label{fig:agtb}  
\end{figure}

\begin{figure}[!ht]
\begin{center}  
\epsfig{file=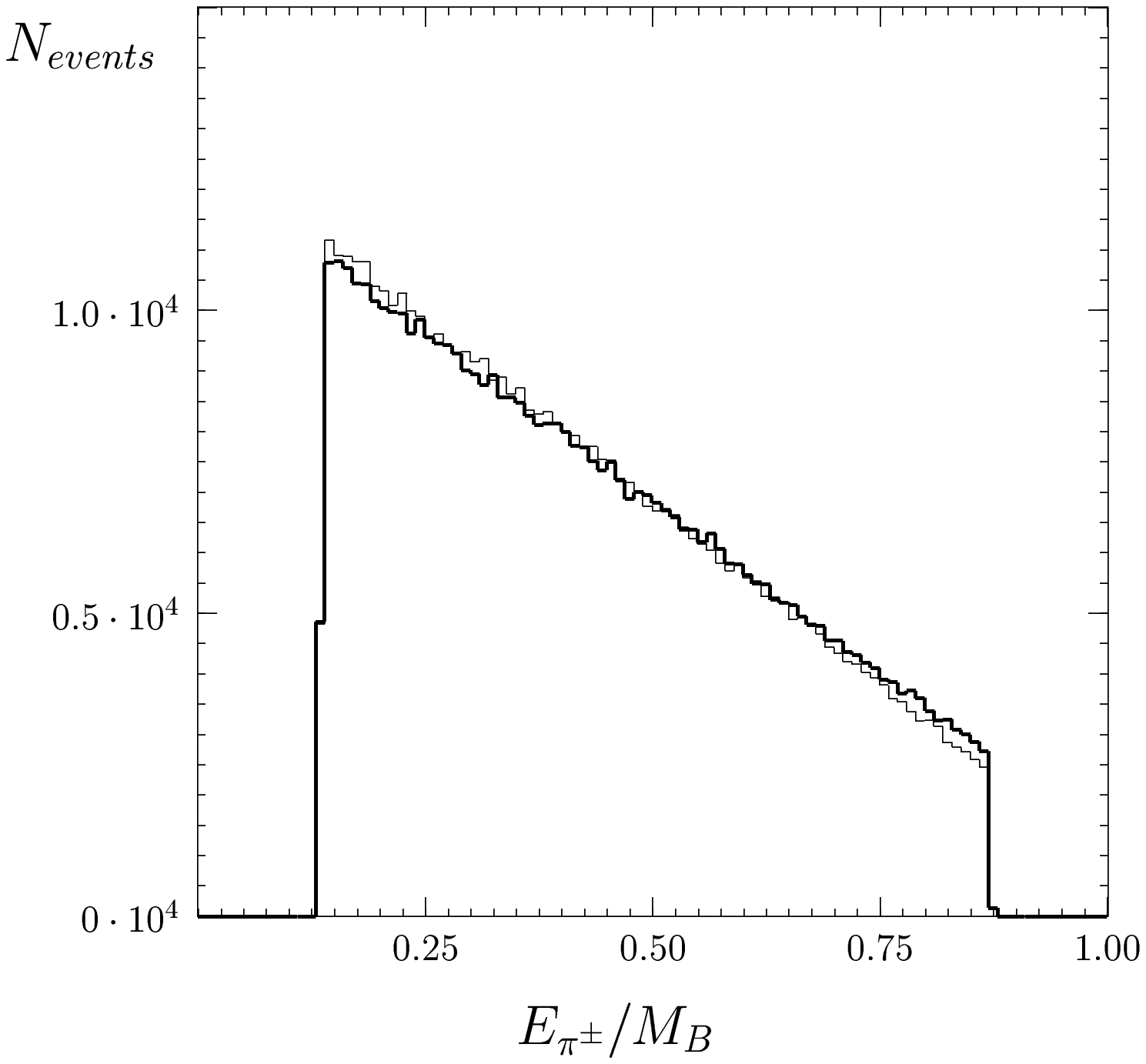,width=70mm,height=60mm}
\epsfig{file=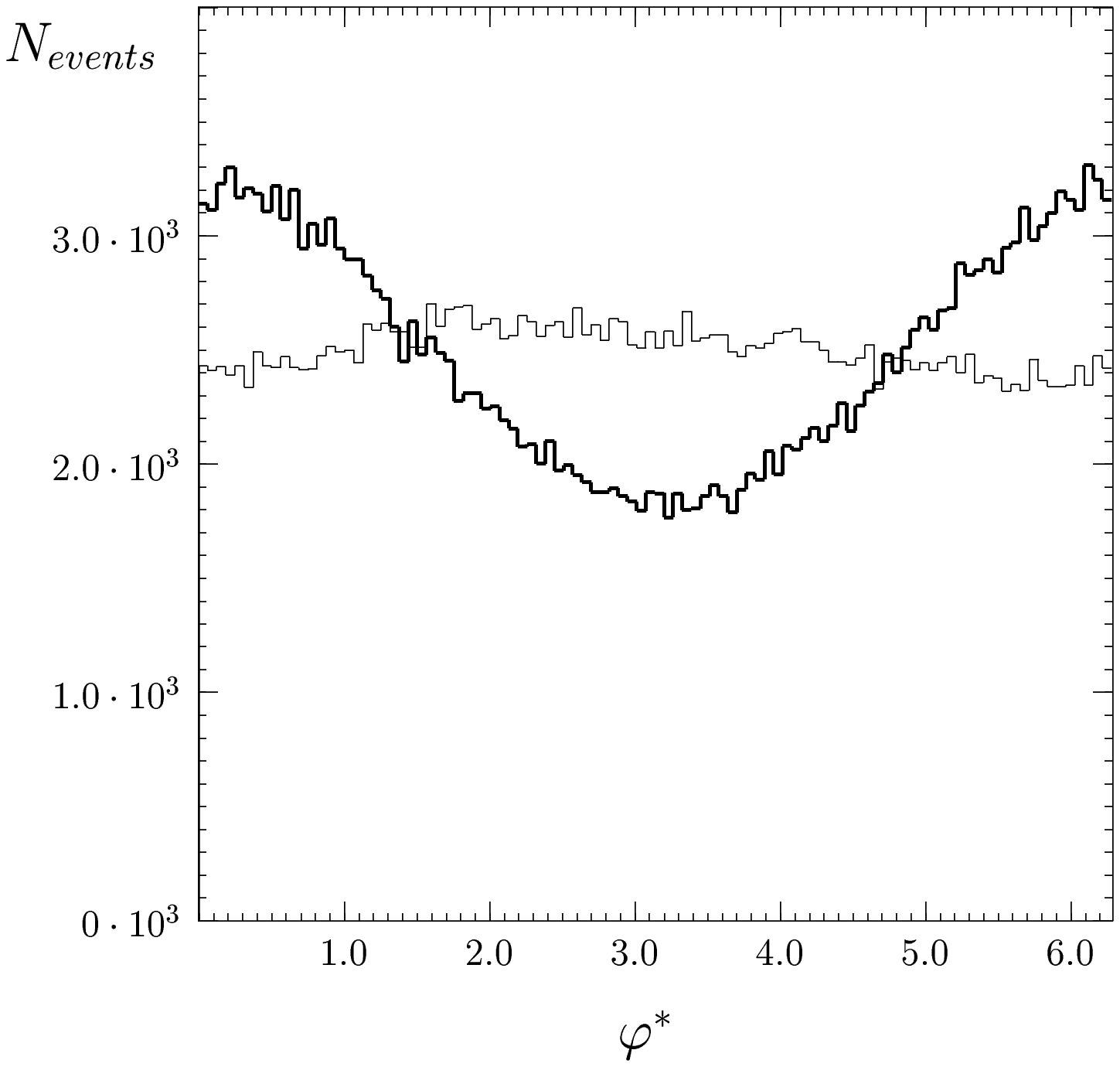,width=70mm,height=60mm}
\end{center} 
\caption 
{\it As in the upper panels of 
figure \ref{fig:delta07} but for $\delta_{\rm CP}=0.7$ 
and $a=0.8\, b$.} 
\label{fig:altb}  
\end{figure}

As we discussed, the relation $a=\pm b$ is only approximate. Therefore, 
in Figure~\ref{fig:agtb} (\ref{fig:altb}) we show the $\pi^\pm$ spectra and 
acoplanarities for $a$ greater (smaller) than $b$, but keeping as previously 
their phases equal (and assuming $|q/p|=1$). We take $b=0.8~a$ in 
Fig.~\ref{fig:agtb} and  $a=0.8~b$ in Fig.~\ref{fig:altb}. In both figures
the single CP violating phase $\delta_{\rm CP}=0.7$. 

Fig~\ref{fig:agtb} shows that for $b=0.8~a$ with the same value 
of $\delta_{\rm CP}$ the CP violating effects
in $\pi^\pm$ energy spectra are enhanced, while in the acoplanarities they 
are only slightly affected compared to the case $a=b$, c.f.\  
the upper right panel of Fig.~\ref{fig:delta07}.  

On the other hand, for $a$ approaching $b\beta$ (for $B\to\tau^+\tau^-$ 
decays $\beta\approx0.74$) the effects of CP violation in
the $\pi^\pm$ energy spectra disappear. This is clearly seen by 
comparing the right panel of Fig.~\ref{fig:altb} with the 
upper left one of Fig.~\ref{fig:delta07}. This agrees with our observations
following Eqs.~(\ref{eqn:lambdas}) and (\ref{eq:CPphase}) and with the
discussion in Section \ref{sec:susy}. In contrast, the acoplanarities shown in 
the right panel of Fig.~\ref{fig:altb} clearly indicate the CP violation
even for $a\approx b\beta$ confirming our discussion in Section
\ref{sec:spindensity}.
This clearly demonstrates the complementarity of the energy and acoplanarity 
distributions as a means to detect CP violation.

\section{Conclusions}

In this letter we have investigated possible signals of CP violation in
the decays $B^0(\bar{B}^0)\rightarrow\tau^+\tau^-$. We have developed the
necessary formalism and numerical tools allowing to apply the {\tt TAUOLA} 
$\tau$-lepton decay library together with its {\tt universal interface} 
to simulate fully the effects of the polarization of $\tau^+$ and $\tau^-$ 
originating from such decays.

We have argued that in the interesting new physics scenario of supersymmetry
with $\tan\beta\sim40\div50$,  in which the rates of 
$B_d^0(\bar{B}_d^0)\rightarrow\tau^+\tau^-$ decays are enhanced and  could
be detectable in the SLAC and KEK $B$-factories, the dependence
of the CP asymmetries on the model parameters simplifies. Moreover,
the CP violating phase needs not be small. In the non-minimal 
flavour violation case it can be of the same order as the phase of the 
$V_{td}$ element of the CKM matrix. Therefore the CP asymmetries
can be quite large as opposed  to the $B^0(\bar{B}^0)\rightarrow\mu^+\mu^-$ 
decays in which they are kinematically suppressed. 

By using Monte-Carlo simulations we have investigated the possible 
effects of CP violation in two realistic  experimental observables 
and demonstrated that they might be detectable if the CP violating phase
is reasonably large, {\it i.e.} ${\cal O}(1)$. 

Since the decays $B^0(\bar{B}^0)\rightarrow\tau^+\tau^-$ have not been
discovered yet, we have not discussed the statistics requirements nor 
attempted at including in our analysis any systematic or detector 
uncertainties. It is clear that once these decays are discovered, other 
$\tau$ decay channels than the ones investigated here can be analysed 
jointly to give additional information on the polarization of $\tau$'s.
Our numerical tools are prepared for that. The tools can also be 
applied to determine the upper limits
on the branching fraction of the $B^0(\bar{B}^0)\rightarrow\tau^+\tau^-$
decays by the \babar\ and BELLE collaborations. 

As a final remark, we point that our analysis can be taken over 
to Higgs boson production at linear colliders with its subsequent
decay to $\tau$ pairs, which as yet has not been simulated in connection
with the complex scalar and pseudoscalar couplings.

\vskip0.3cm
\noindent {\bf Acknowledgements}\\
The authors would like to thank P. Ball, A. Martin, M.~Misiak,  
H. Pa\l ka and J. Stirling for useful discussions.  
P.H.Ch. thanks the CERN Theory Group 
for hospitality during the initial stage of this research. 

Work supported by the Polish State Committee for Scientific 
Research Grants 2 P03B 040 24 for years 2003-2005 (P.H.Ch. and J.K.), 
1 P03 091 27 for years 2004-2006 (Z.W.), 1 P03B 009 27  for  years 2004-2005
(M.W.), and by the EC Contracts 
HPRN-CT-2000-00148 (P.H.Ch.)  and HPRN-CT-2000-00149 (J.K).  In addition, 
M.W. acknowledges the Maria Sk\l odowska-Curie Fellowship granted by the 
the European Community in the framework of the Human Potential Programme 
under contract HPMD-CT-2001-00105 
({\it ``Multi-particle production and higher order correction''}).

\end{document}